\documentclass[prd,a4paper,nofootinbib]{revtex4}

\newcommand{\nn}{\nonumber}
\newcommand{\beq}{\begin{equation}}
\newcommand{\eeq}{\end{equation}}
\newcommand{\bea}{\begin{eqnarray}}
\newcommand{\eea}{\end{eqnarray}}
\newcommand{\ba}{\begin{array}}
\newcommand{\ea}{\end{array}}
\newcommand{\bec}{\begin{center}}
\newcommand{\eec}{\end{center}}
\newcommand{\bei}{\begin{itemize}}
\newcommand{\eei}{\end{itemize}}
\newcommand{\AddrAHEP}{
  AHEP Group, Instituto de F\'{\i}sica Corpuscular --
  C.S.I.C./Universitat de Val{\`e}ncia \\
  Edificio Institutos de Paterna, Apt 22085, E--46071 Valencia, Spain}

\usepackage{pstricks}
\usepackage{graphicx}
\usepackage{inputenc}
\usepackage{pst-coil}
\usepackage{longtable}
\usepackage{amssymb}
\usepackage{amsfonts}
\usepackage{bbm}
\usepackage{color}

\def\vev#1{\left\langle #1\right\rangle}

\def\hbar{\hspace{0pt}\raisebox{1pt}{$-$} \hspace{-7pt} h}

\def\5{\overline 5}

\begin{document}

\begin{flushright}
SACLAY-T08/126\\
IFIC/08-43\\
\end{flushright}
\vspace{0.3cm}

\title{Fermion masses and mixing in models with $SO(10)\times A_4$ symmetry}

\author{Federica Bazzocchi$^1$, Michele Frigerio$^2$ and  Stefano Morisi$^1$\\
\vskip 10pt
{\small\it $^1$\AddrAHEP\\
$^2$ Institut de Physique Th\'{e}orique, CEA-Saclay, 
91191 Gif-sur-Yvette Cedex, France}}

\begin{abstract}
We study the flavour sector in models where the three families of matter are
unified in a $(16,3)$ representation of the $SO(10)\times A_4$ group.
The necessary ingredients to realize tri-bi-maximal mixing in the lepton
sector are identified systematically.
The non-renormalizable operators contributing to the fermion mass
matrices play an important role.
We also present a mechanism to explain the 
inter-family mass hierarchy of quarks and charged leptons,
which relies on a `universal seesaw' mechanism
and is compatible with tri-bi-maximal mixing.
\end{abstract}

\maketitle

\section{Introduction}

Before the experimental determination of neutrino oscillation parameters, a common theoretical prejudice was the
expectation of small lepton mixing angles, by analogy with the CKM angles in the quark sector. This was thought to be a generic prediction of
Grand Unification Theories, as they incorporate some form of quark-lepton symmetry. On the contrary, experiments have
shown that the $2-3$ lepton mixing is close to maximal, the $1-2$ mixing is large, while the $1-3$ angle can be as
large as the Cabibbo angle \cite{Maltoni:2004ei,Fogli:2005cq}. With the benefit of hindsight, this disparity
between quarks and leptons was interpreted as a manifestation of the Majorana nature of neutrinos. From then on, many grand
unified models have been built, 
in which large lepton mixing angles appear naturally (see e.g.
\cite{BB,ABB,BPW,BSV,GMN}).

Several conjectures on special  `symmetric' values of the lepton mixing angles have
been also proposed. For example, the
bi-maximal mixing pattern ($\theta_{23}=\theta_{12}=\pi/4$ and $\theta_{13}=0$) was extensively studied but it is by now
excluded, since solar neutrino experiments require $\theta_{12} = (35 \pm 4)^\circ$ at $3$-$\sigma$. Current data are in
very good agreement, instead, with the so-called tri-bi-maximal (TBM) mixing scheme
\cite{Harrison:2002er}, where the
second and third neutrino mass eigenstates have the following flavour content:
$\nu_2= (\nu_e+\nu_\mu+\nu_\tau)/\sqrt{3}$
and $\nu_3=(\nu_\mu-\nu_\tau)/\sqrt{2}$. Such special values of the angles may point to a non-abelian family symmetry. In
particular, the TBM scheme seems to require that the three families of lepton doublets belong to the same dimension three
representation, the simplest suitable symmetry being the discrete group $A_4$ 
\cite{Ma:2001dn,Babu:2002dz} (early applications of $A_4$ as a family symmetry
are discussed in \cite{wyler,BNR}).

While generically large lepton mixing angles found their place in the framework of Grand Unification, it is a much tougher
task to justify special values of such angles. The first few attempts to build unified models with TBM mixing have
encountered several difficulties and concrete realizations so far require many degrees of complication. In most $A_4$
models for TBM mixing, the lepton doublets and singlets transform differently under the family symmetry, so that the
allowed embedding is $SU(5)$ unification \cite{Ma:2006sk,Chen:2007afa,Altarelli:2008bg}, or the Pati-Salam group
\cite{King:2006np}. The alternative option, to take all leptons transforming as triplets under $A_4$, was later considered
\cite{Chen:2005jm} and it was shown \cite{Ma:2006wm} to be suitable to achieve TBM mixing. This assignment is compatible
with $SO(10)$ unification of the gauge interactions. Few models based on the $SO(10)\times A_4$ group have been studied 
\cite{Morisi:2007ft,Grimus:2008tm}. A different scheme to achieve TBM mixing
in $SO(10)$ unification makes use of an
$SU(3)$ family symmetry \cite{deMedeirosVarzielas:2005ax} or its discrete subgroup 
$\Delta(27)$ \cite{deMedeirosVarzielas:2006fc}.

In this paper, we investigate systematically the structure of fermion masses and mixing in models with $SO(10)$
unification and $A_4$ family symmetry. This is the minimal framework for a complete unification of the three families of
matter, since the three $SO(10)$ spinors of dimension 16 can transform as a triplet under $A_4$. The scheme for realizing
TBM lepton mixing consists of breaking $A_4$ into its $Z_3$ and $Z_2$ subgroups in the charged lepton and neutrino sectors,
respectively. This misalignment results in TBM values for the lepton mixing angles. On the other hand, the flavour
alignment of the up and down quark sectors should be approximately maintained, in order to explain the smallness of the
CKM angles. Since in our framework all matter fields belong to the same representation of $SO(10)\times A_4$, the Yukawa
couplings of the different sectors are strictly related and it is particularly
challenging to satisfy all constraints.

More specifically, one faces the need to generate two independent structures for the 
up-quark mass matrix $M_u$ and the
neutrino Dirac mass matrix $M_\nu$, which in $SO(10)$ are strictly related. It has been shown \cite{Morisi:2007ft} that a
sufficiently complicated arrangement of higher dimensional operators can disentangle the two structures. Another way out \cite{Grimus:2008tm} 
is to assume that the light neutrino mass matrix, $m_\nu$, 
is independent from the form of $M_\nu$ and it is 
instead generated 
by the coupling of two lepton doublets to a Higgs triplet. 
However, in the context of $SO(10)$ theories this
assumption cannot be realized exactly, as discussed later. 
Another difficulty of $A_4$ models, which 
becomes even more severe in $SO(10)$, is the
need to introduce non-vanishing CKM parameters without generating too large deviations from the TBM lepton mixing. We
will reconsider these issues, 
generalizing previous results and identifying new solutions.

Perhaps the most serious shortcoming of $A_4$ flavour models is that the hierarchy between the masses of the three
families can be accommodated, but  is not explained. This problem can be addressed \cite{Altarelli:2005yp} in $A_4$
models where lepton doublets transform as $L_i\sim 3$ while charged lepton singlets transform as $e^c_i\sim 1,1',1''$.
In this case one can introduce an extra family symmetry that distinguishes the three families. For example, a $U(1)$ family
symmetry of the Froggatt-Nielsen type \cite{Froggatt:1978nt}, with different charges for the three $e^c_i$,
can explain the hierarchy of the charged lepton masses. This mechanism can be promptly extended in an $SU(5)$-invariant
fashion to the down and up quarks. The origin of the inter-family hierarchy
is more problematic in $SO(10)$, since
both chiralities of matter transform as $A_4$ triplets. As a consequence, a universal mass term of the type $m
\psi_i\psi^c_i$ is allowed \cite{DMVshort} and no extra symmetry can distinguish 
the families. We will argue that a natural
solution for this problem emerges from the mixing with heavy vector-like families. The hierarchy arises in the quark and
charged lepton sectors precisely because there $A_4$ is broken to $Z_3$ and each family transforms differently under this
residual symmetry.

The paper is organized as follows. In section \ref{route} (and in appendix \ref{B1}) we recall the structure of the
fermion mass matrices which lead to TBM mixing in $A_4$ models. In section \ref{sec:ren} (and in appendix \ref{A1}) we
study $SO(10)\times A_4$ models with renormalizable Yukawa couplings and
investigate the possibility to
realize the TBM mixing scheme. In section \ref{sec:nonr} (and in appendix \ref{A2}) we extend our analysis to dimension five
operators, which overcomes several difficulties of the renormalizable models. In section \ref{sec:3h} we present
the mechanism to explain the mass hierarchy between the three families, in a way
that is compatible with TBM mixing. In section
\ref{sec:con} we summarize our main results.

\section{The route to tri-bi-maximal mixing \label{route}}

In this section we review a scheme to achieve TBM lepton mixing,
that was employed in the construction of several $A_4$ models 
\cite{Ma:2004zv,
Altarelli:2005yp,
Babu:2005se,
Ma:2005mw,
Zee:2005ut,
Altarelli:2005yx,
He:2006dk,
Altarelli:2006kg,
Ma:2006vq,
Bazzocchi:2007na,
Lavoura:2007dw,
Brahmachari:2008fn,
Hirsch:2008rp,
Lin:2008aj,
Feruglio:2008ht} 
and turns out to be the one suitable in the context of $SO(10)$ unification \cite{Ma:2006wm, Morisi:2007ft,Bazzocchi:2007au,Grimus:2008tm}.

Let us introduce a set of simple mass matrix structures 
which can accommodate all fermion masses, leading at the same time to TBM lepton mixing and vanishing quark mixing.
We adopt a supersymmetric notation, with 
$-{\cal L}_m= uM_uu^c+dM_dd^c+eM_ee^c+\frac 12 \nu m_\nu \nu$, where
\begin{equation}
\label{mass}
M_f =
\left( \begin{array}{ccc} A_f& B_f & C_f\\ C_f & A_f & B_f\\ B_f& C_f &A_f   \end{array}\right) ~,~~~~f=u,d,e,\qquad
\qquad m_\nu =
\left(
\begin{array}{ccc}
a& 0&b\\
0& c &0\\
b& 0 &a
\end{array}
\right)\,.
\end{equation}
These charged fermion and neutrino Majorana mass  matrices can be diagonalized as
follows: 
\beq\ba{l}
D_f = U_f^T M_f U_f^* =  diag(\,A_f+B_f+C_f,\; 
A_f+\omega^2 B_f + \omega C_f,\; A_f+\omega B_f + \omega^2 C_f\,) 
\equiv diag(m_{f1},m_{f2},m_{f3}) ,\\
d_\nu  = U_\nu^T m_\nu U_\nu = 
diag(\,a+b,\; c,\; a-b\,)\equiv  diag(m_1,m_2,m_3),
\ea\label{diag}\eeq
where $\omega \equiv e^{2\pi i/3} = (-1 +\sqrt{3} i)/2$
and the unitary mixing matrices are given by
\begin{equation}
U_f = U_\omega \equiv \frac{1}{\sqrt{3}}\left(
\begin{array}{ccc}
1&1&1\\
1&\omega&\omega^2\\
1&\omega^2&\omega
\end{array}
\right)~,\qquad U_\nu=
\frac{1}{\sqrt{2}}\left(
\begin{array}{ccc}
1 & 0 & -1 \\
0 & \sqrt{2} & 0 \\
1 & 0 & 1 
\end{array}
\right)
\,.
\label{ufn}\end{equation}
The matrix $U_\omega$ is known as `magic' mixing matrix. We will refer to the
matrix structures in Eq.~(\ref{mass}) as `magic' and `cross' mass matrix, respectively.

There is no observable quark mixing, since
$U_{CKM} \equiv U_u^\dag U_d = U_\omega^\dag U_\omega = {\mathbbm 1}_3$.
On the contrary, the leptons mix tri-bi-maximally:
\beq
U_{lepton} \equiv U_e^\dag U_\nu  = U_\omega^\dag U_\nu =
diag(1,\omega^2,\omega) \; U_{TBM} \;diag(1,1,-i) ~,\qquad
U_{TBM} \equiv
\left(\ba{ccc}
\sqrt{\frac 23} & \sqrt{\frac 13} & 0 \\
-\sqrt{\frac 16} & \sqrt{\frac 13} & \sqrt{\frac 12} \\
-\sqrt{\frac 16} & \sqrt{\frac 13} & -\sqrt{\frac 12} \\
\ea\right) ~.
\eeq
The diagonal phase matrices can be absorbed by rephasing
the charged lepton fields and $m_3$.
Notice that several different choices of the pair of unitary matrices 
$U_e$ and $U_\nu$ may reproduce TBM mixing.
However, as shown in Appendix \ref{B1}, 
the choice made in Eq.~(\ref{ufn}) is the only one justified by the $A_4$ family
symmetry.

In ordinary $SO(10)$ models, all light matter fields reside in three
dim-16 representations, which also contain right-handed neutrinos.
Then, the light Majorana neutrino mass matrix is given in general by
\beq
m_\nu = M_L - M_\nu M_R^{-1} M_\nu^T ~,
\label{seesaw}\eeq
where $M_L$ is a direct Majorana mass term for the left-handed neutrinos,
$M_\nu$ is the Dirac-type neutrino mass matrix and
$M_R$ is the Majorana mass matrix of right-handed neutrinos.
The second term in Eq.~(\ref{seesaw}) 
is the outcome of the type I seesaw mechanism \cite{mink,GRS,yana,gla,mose},
while the first term may arise from a type II seesaw \cite{MW,SV,LSW,mose2}.
The latter can be sizable in models with at least one $\overline{126}$ Higgs multiplet
and a scalar potential such that the $SU(2)_L$ triplet component of
$\overline{126}$ develops a vacuum expectation value (VEV). In all other cases the 
type I seesaw dominates and then the cross structure of $m_\nu$ 
in Eq.~(\ref{mass})
should emerge from the interplay of the $M_\nu$ and $M_R$ structures.
The simplest possibility is, of course,
\beq
M_\nu =
\left( \begin{array}{ccc} A_\nu& 0 & B_\nu\\ 0 &  C_\nu & 0\\ B_\nu & 0 & A_\nu
\end{array}\right) ~,~~~
\qquad M_R =
\left(
\begin{array}{ccc}
A_R&0&B_R\\
0&C_R&0\\
B_R&0&A_R
\end{array}
\right)\,,
\label{seesawI}
\eeq
with at least one among $B_\nu$ and $B_R$ being non-zero. 
Another possibility which may be justified by the $A_4$ family symmetry is
given by
\beq
M_\nu =
\left( \begin{array}{ccc} 0 & A_\nu & 0 \\ -A_\nu &  0 & A_\nu \\ 0 & -A_\nu & 0
\end{array}\right) ~,
\label{antinu}\eeq
with $M_R$ as in Eq.~(\ref{seesawI}).
In this case one recovers the form of $m_\nu$ in Eq.~(\ref{mass}) with
$b=-a$, which implies a normal neutrino mass hierarchy.

\section{Models with only renormalizable Yukawa couplings}
\label{sec:ren}

In this section we consider the Yukawa couplings $Y \Psi \Psi \Phi$
invariant under $SO(10)\times A_4$ and investigate whether 
they can generate the mass matrix structures in Eq.~(\ref{mass}) and thus realize
TBM lepton mixing.

We will indicate with $(R,\tilde{R})$ a multiplet transforming
in the representation $R$ of $SO(10)$ and $\tilde{R}$ of $A_4$. Matter fields are
unified into a single multiplet $\Psi\sim(16,3)$.
The structure of the mass matrices depends on the assignment of 
the Higgs multiplets $\Phi$ under $SO(10)\times A_4$ and on the components of
$\Phi$ which acquire a non-zero VEV.
We will indicate with 
$\langle r^\Phi \rangle \equiv v_r^\Phi$ the VEV of the $\Phi$ component 
in the representation $r$ of $SU(5)$.  
All the possible matrix structures are listed in Appendix \ref{A1}.

Let us discuss first the mass matrices of charged fermions.
As follows from Eq.~(\ref{diag}), the masses of the three families can
be accommodated only if the matrix entries $A_f$, $B_f$ and $C_f$ are all non-zero,
of the same order of magnitude, and different from each other. 
One can generate the up-quark mass matrix $M_u$
introducing the Higgs multiplets $\phi\sim (10,1)$, $\eta\sim(10,3)$ and 
$\rho\sim(120,3)$. The VEVs
\beq
\langle 5^\phi \rangle = v_5^\phi ~,~~~~~
\langle 5^{\eta}_i \rangle = (v_5^{\eta},v_5^{\eta},
v_5^{\eta}) ~,~~~~~
\langle 45^\rho_i \rangle = (v_{45}^\rho,v_{45}^\rho,v_{45}^\rho) 
\label{vu}\eeq
generate $A_u$, $B_u+C_u$ and $B_u-C_u$, respectively.
Alternatively, $\phi$ and/or $\eta$ can be replaced by 
$\Delta \sim (\overline{126},1)$ and/or $\Omega \sim (\overline{126},3)$
with the same VEV alignment.

In the down quark and charged lepton sector,
there are two classes of contributions to the mass matrices:
those of the type $\delta M_e^T=\delta M_d$
can be generated by  the VEVs
\beq\label{vevd1}
\langle \overline{5}^\phi \rangle = v^\phi_{\overline{5}} ~,~~~~
\langle \overline{5}_i^\eta \rangle = (v^\eta_{\overline{5}},v^\eta_{\overline{5}},
v^\eta_{\overline{5}}) ~,~~~~
\langle \overline{5}_i^\rho \rangle = (v^\rho_{\overline{5}},
v^\rho_{\overline{5}},v^\rho_{\overline{5}}) ~;
\label{vd5}\eeq
those of the type $\delta M_e^T = -3 \, \delta M_d$
can be generated by the VEVs
\beq\label{vevd2}
\langle \overline{45}^\Delta \rangle = v_{\overline{45}}^\Delta ~,~~~~
\langle \overline{45}_i^{\Omega} \rangle = (v^\Omega_{\overline{45}},
v^\Omega_{\overline{45}},v^\Omega_{\overline{45}}) ~,~~~~
\langle \overline{45}_i^{\rho} \rangle = (v^\rho_{\overline{45}}
v^\rho_{\overline{45}},v^\rho_{\overline{45}}) ~.
\label{vd45}\eeq
The first, second and third term in Eqs.~(\ref{vd5}) and (\ref{vd45})
contribute to $A_{d,e}$, $B_{d,e}+C_{d,e}$ and $B_{d,e}-C_{d,e}$, respectively.
When all six contributions are present,
the three masses of down quarks and charged leptons can be fitted.
If some of the VEVs are zero, non-trivial
relations between the masses are predicted.
For example, an economical scenario with no VEVs in the Higgs doublets of the
$\overline{126}$ multiplets would imply $A_e=A_d$
and $B_e+C_e=B_d+C_d$. In this limit one predicts (i)
$|e^{i\varphi_\mu}m_\mu+m_\tau|=|e^{i\varphi_s} m_s+m_b|$ with
$\varphi_{\mu,s}$ arbitrary, which may be compatible with the GUT scale values
(see e.g. \cite{RoSe}), and (ii)
$m_e=m_d$, which requires corrections of the order of one MeV.

Let us now move to the neutrino sector.
Consider first the term $M_L$ in Eq.~(\ref{seesaw}).
It is generated when
the $SU(2)_L$ Higgs triplets in $\overline{126}$ multiplets
receive a tiny VEV through a type II seesaw mechanism
\cite{MW,SV,LSW,mose2}. Taking
\beq
\langle 15^{\Delta} \rangle = v_{15}^{\Delta} ~,~~~~
\langle 15_i^{\Omega} \rangle = (0,v_{15}^{\Omega},0) ~,
\label{trip}\eeq
$M_L$ acquires the structure of $m_\nu$ in Eq.~(\ref{mass}) 
with $a=c$. Therefore, if one assumes $m_\nu=M_L$,
the TBM lepton mixing is realized
with one constraint on the light neutrino mass spectrum
\cite{Altarelli:2005yp}.

One should stress, however, that 
the type I seesaw contribution 
(second term in Eq.~(\ref{seesaw})) is necessarily present in $SO(10)$ models
and it turns out to be incompatible with the TBM structure.
In fact, the VEVs $v_5^\phi$ and $v_5^\eta$, which are necessary to generate
$M_u$, also contribute to the neutrino Dirac mass matrix $M_\nu$, which takes
the form
\beq
M_\nu = \left(\ba{ccc} 
A_\nu & B_\nu & B_\nu \\
B_\nu & A_\nu & B_\nu \\
B_\nu & B_\nu & A_\nu \\
\ea\right)~,
\eeq  
with $A_\nu=A_u$ and $B_\nu = (B_u+C_u)/2$.
The right-handed neutrino mass matrix $M_R$ is generated by the VEVs
of the $SU(5)$ singlet components of the $\overline{126}$ Higgs multiplets.
By constructing the combination $M_\nu M_R M_\nu^T$ for all the possible $A_4$ structures of $M_R$ (see Table \ref{tab:fl4D})
and barring fine-tuning of independent
couplings, 
we find that the term  $B_\nu\propto v^\eta_5$ is never compatible with the
structure of $m_\nu$ given in Eq.~(\ref{mass}). 
This shows that the exact realization of the TBM mixing is not possible.

In order to estimate the deviation from TBM mixing,
a reasonable hypothesis is that $M_R$ acquires a structure analog to $M_L$
from the VEVs  
\beq
\langle 1^\Delta \rangle = v_1^\Delta ~,~~~~ 
\langle 1_i^\Omega \rangle = (0,v_1^\Omega,0) ~.
\label{vevR}\eeq
In this case $M_R$ has the form in Eq.~(\ref{seesawI}) with $A_R=C_R$ and
one finds
\beq\label{type12}
m_\nu= \left(\ba{ccc}
a&0&b\\
0&c&0\\
b&0&a
\ea\right) + \left(\ba{ccc}
0&\epsilon&\epsilon\\
\epsilon&0&\epsilon\\
\epsilon&\epsilon&0
\ea\right)~.
\eeq
The effect of type I seesaw is to make $a\ne c$ and to add the $\epsilon$ term.
It is easy to check that,  in the basis in which the  charged leptons are diagonal, 
$m_\nu$ is still $\mu-\tau$ symmetric, therefore the values of the mixing angles 
$\theta_{13}=0$ and $\theta_{23}=\pi/4$ are preserved.
On the contrary, 
the TBM value of $\theta_{12}$ and the light neutrino mass spectrum are corrected in
a correlated way.
Taking for simplicity all parameters real, one finds
\beq\label{s12}
\tan 2\theta_{12}=\frac{2\sqrt{2}}{1 + \displaystyle{\frac{9\epsilon}{a+b-c}}}~. 
\eeq
The TBM value of $\theta_{12}$ is recovered for $\epsilon=0$.
In order to accommodate the experimental value of $\theta_{12}$ at the 
2-$\sigma$ level \cite{Maltoni:2004ei}, one needs    $-0.03< \epsilon/(a+b-c) < 0.04$,
so that a few percent type I seesaw contribution can be tolerated. 
In the case $v_1^\Omega=0$, the same discussion holds with  $a=c$.

We have shown that, if  $M_u$ is given by Eq.~(\ref{mass}),
then the structure of $M_\nu$ is not compatible with exact TBM mixing.
Vice versa, it is instructive to consider the structures of $M_\nu$ leading to TBM mixing
and see what are the implications for the up quark sector.
A first possibility is to take the $M_R$ structure of Eq.~(\ref{seesawI}),
generated by the VEVs in Eq.~(\ref{vevR}), and the $M_\nu$ structure of Eq.~(\ref{antinu})
by the VEV
\beq
\langle 5^{\rho}_i\rangle = (v_5^\rho,0,v_5^\rho)~.
\eeq
Even though this
VEV alignment does not preserve any subgroup of $A_4$, we expect that it may
be justified dynamically; the analog alignment has been obtained e.g.
in a model with $SU(3)$ family symmetry \cite{KR}. 
This would be sufficient to achieve TBM mixing in the lepton sector.
In the context of $SO(10)$, however, 
the only VEV that may contribute to $M_u$ without modifying the structure
of  $M_\nu$ in Eq.~(\ref{antinu})
is $v_{45}^\rho$ (see table \ref{tab:ga4D}).
Therefore  $M_u$ would be purely
antisymmetric ($A_u=0$ and $B_u=-C_u$) and thus 
would have one zero and two equal eigenvalues, which clearly is not acceptable.
A second possibility to realize exact TBM mixing
is to generate the structure of $M_\nu$ in Eq.~(\ref{seesawI})
by modifying  the VEV alignment of $5^\eta_i$ in Eq.~(\ref{vu}) as follows:
\beq
\langle 5_i^\eta \rangle = (0,v_5^\eta,0)~. 
\eeq
In this case the up quark mass matrix takes the form
\beq
\label{nomagic}
M_u = \left(\ba{ccc} A_u & (B_u-C_u)/2 & C_u \\ 
(C_u-B_u)/2 & A_u & (B_u-C_u)/2 \\ B_u & (C_u-B_u)/2 & A_u
\ea\right)~.
\label{mrn}\eeq
Since $M_d$ has the structure in Eq.~(\ref{mass}),
the requirement to have $V_{CKM}\approx {\mathbbm 1}_3$ 
would imply $C_u\approx -B_u$ 
and thus lead to the wrong relation $m_c\approx m_t$
(see Eq.(\ref{diag})). Vice versa, the requirement
to fit the three up quark masses would force  the $V_{CKM}$ angles to be large. 
We conclude that the strong departures from the $M_u$ structure in Eq.~(\ref{mass}),
required to achieve exact TBM lepton mixing,
are not viable phenomenologically.

A comment is in order on the pattern of $SO(10)\times A_4$ spontaneous breaking. 
The triplet VEVs of the type $(1,1,1)$, introduced in Eqs.~(\ref{vu})-(\ref{vd45}), 
break $A_4$ to a $Z_3$ subgroup at the electroweak scale. 
The VEV of the type $(0,1,0)$, introduced in Eq.~(\ref{trip}), breaks
$A_4$ to a $Z_2$ subgroup at the scale of light neutrino masses
(notice that the large $A_4$ breaking VEV $v_1^\Omega$ in Eq.~(\ref{vevR}) is
not needed in a minimal scenario).
This misalignment between the charged fermion sector and the neutrino sector is a crucial ingredient to explain TBM mixing. 
On the $SO(10)$ side, one faces the problem to generate
VEVs only in some specific components of the Higgs multiplets, 
specifically the Higgs doublets in $10$ and $120$ multiplets and the
Higgs triplets and singlets in $\overline{126}$ multiplets.
It is a difficult task to arrange for an appropriate scalar potential,
also in view of the different energy scales of different sets of VEVs. 
The analysis of such potential, which is beyond the scope of this paper, 
may in principle reveal a connection between the VEV alignment dynamics
in the $A_4$ and $SO(10)$ sectors.

In the rest of this section, we will show that a relatively economic model can be built
by using only $SO(10)$ Higgs multiplets which are singlet under $A_4$, plus
a set of gauge singlet flavon fields responsible for the breaking of the family symmetry.
In this approach the problem of VEV alignments can be treated separately
in the $SO(10)$ and $A_4$ sectors. Another advantage is that the number
of $SO(10)$ multiplets can be considerably reduced, which is desirable to maintain
the theory perturbative well above the GUT scale.

\begin{table}[t]
\begin{center}
\begin{tabular}{|c|c|ccc|cccc|}
\hline
& matter & \multicolumn{3}{|c|}{Higgs fields} & \multicolumn{4}{|c|}{flavons} \\
\cline{2-9}
& $\Psi$ & $\phi$ &  $\rho$ & $\Delta$ & $\sigma$ & $\chi$ & $\tau$ & $\varphi$ \\
\cline{2-9}
$SO(10)$ & 16 & 10 & 120 & $\overline{126}$ & 1 & 1 & 1 & 1 \\
$A_4$ &3&1&1&1&1&3&1&3\\
$Z_N$ &1&$\alpha^m$&$\alpha^m$&$\alpha^r$&
$\alpha^{-m}$&$\alpha^{-m}$&$\alpha^{-r}$&$\alpha^{-r}$ \\
\hline
\end{tabular}
\end{center}
\caption{Chiral superfields in a minimal renormalizable model for 
approximate TBM mixing. 
Here $\alpha^N=1$ and $1\le m\ne r < N$.}
\label{rent}
\end{table}

The field content of the model is given in Table \ref{rent}.
In order to couple the appropriate flavon fields only to certain $SO(10)$ operators,
we introduced a $Z_N$ symmetry, the minimal choice being $N = 3$
with charges $m=1$ and $r=2$.
The Yukawa superpotential is then given by
\beq
W_Y =\frac{y_1 \sigma  + y_3 \chi}{\Lambda}\Psi\Psi\phi  
+ \frac{g_3 \chi}{\Lambda}\Psi\Psi\rho
+\frac{f_1 \tau  + f_3 \varphi}{\Lambda}\Psi\Psi \Delta ~,
\label{supren}\eeq
where $\Lambda$ is a large energy scale where the $A_4\times
Z_N$ symmetry is realized.
The flavon fields are assumed to acquire the VEVs
\beq
\langle \sigma \rangle = \sigma ~,\quad
\langle \chi_i \rangle = (\chi,\chi,\chi) ~,\quad
\langle \tau \rangle = \tau ~,\quad
\langle \varphi_i \rangle = (0,\varphi,0) ~,
\label{flavev}\eeq
which break $A_4$ as well as $Z_N$ at some scale $\sim \lambda \Lambda$,
with $\lambda < 1$.
The problem to achieve dynamically such VEV alignment has been addressed and
solved in several papers \cite{Altarelli:2005yp,Babu:2005se,Altarelli:2005yx,DMVshort}.
Notice that in Eq.~(\ref{supren}) we included only the leading order operators, linear
in the flavon fields. Higher dimensional operators with $1+n$ flavons 
generate corrections to the mass matrices of relative order $\lambda^n$, which
may be significant for small $n$ and $\lambda$ close to one. However, 
all the operators up to some given $n$ can be forbidden by using a $Z_N$ with a sufficiently large $N$ and
choosing carefully the charges $m$ and $r$ in Table \ref{rent}.

We suppose that all the Higgs doublet in $\phi$ and $\rho$ acquire a VEV at the electroweak scale, except the one in $5^\rho$. 
Also, $1^\Delta$ acquires a VEV at the GUT scale
and the Higgs triplet in $15^\Delta$ takes a VEV of the order of the
light neutrino mass scale. Then, the mass matrices of charged fermions
are explicitly given by 
\bea
M_u &=&  \frac{1}{\Lambda} \left( \begin{array}{ccccc} 
y_1\sigma v_5^{\phi}  & &
y_3\chi  v_5^\phi + g_3 \chi v_{45}^{\rho} & & 
y_3\chi  v_5^\phi - g_3 \chi v_{45}^{\rho} \\ 
y_3\chi  v_5^\phi - g_3 \chi v_{45}^{\rho} & &
y_1\sigma v_5^{\phi}  & &
y_3\chi  v_5^\phi + g_3 \chi v_{45}^{\rho} \\
y_3\chi  v_5^\phi + g_3 \chi v_{45}^{\rho} & & 
y_3\chi  v_5^\phi - g_3 \chi v_{45}^{\rho} & &
y_1\sigma v_5^{\phi}    
\ea\right) ~, \nn\\ 
M_d &=& \frac{1}{\Lambda} \left( \begin{array}{ccccc} 
y_1\sigma v_{\overline{5}}^{\phi}  && 
y_3\chi  v_{\overline{5}}^{\phi} + g_3 \chi (v_{\overline{5}}^\rho+v_{\overline{45}}^\rho) && 
y_3\chi  v_{\overline{5}}^{\phi} - g_3 \chi (v_{\overline{5}}^\rho+v_{\overline{45}}^\rho) \\ 
y_3\chi  v_{\overline{5}}^{\phi} - g_3 \chi (v_{\overline{5}}^\rho+v_{\overline{45}}^\rho) &&
y_1\sigma v_{\overline{5}}^{\phi}  &&
y_3\chi  v_{\overline{5}}^{\phi} + g_3 \chi (v_{\overline{5}}^\rho+v_{\overline{45}}^\rho) \\
y_3\chi  v_{\overline{5}}^{\phi} + g_3 \chi (v_{\overline{5}}^\rho+v_{\overline{45}}^\rho) && 
y_3\chi  v_{\overline{5}}^{\phi} - g_3 \chi (v_{\overline{5}}^\rho+v_{\overline{45}}^\rho) &&
y_1\sigma v_{\overline{5}}^{\phi}    
\end{array}\right) ~, \nn\\
M_e &=& \frac{1}{\Lambda} \left( \begin{array}{ccccc} 
y_1\sigma v_{\overline{5}}^{\phi}  && 
y_3\chi  v_{\overline{5}}^{\phi} -g_3 \chi (v_{\overline{5}}^\rho- 3 v_{\overline{45}}^\rho) && 
y_3\chi  v_{\overline{5}}^{\phi} +g_3 \chi (v_{\overline{5}}^\rho- 3 v_{\overline{45}}^\rho)\\ 
y_3\chi  v_{\overline{5}}^{\phi} +g_3 \chi (v_{\overline{5}}^\rho- 3 v_{\overline{45}}^\rho) &&
y_1\sigma v_{\overline{5}}^{\phi}  &&
y_3\chi  v_{\overline{5}}^{\phi} -g_3 \chi (v_{\overline{5}}^\rho-3v_{\overline{45}}^\rho) \\
y_3\chi  v_{\overline{5}}^{\phi} -g_3 \chi (v_{\overline{5}}^\rho-3v_{\overline{45}}^\rho) && 
y_3\chi  v_{\overline{5}}^{\phi} +g_3 \chi (v_{\overline{5}}^\rho-3v_{\overline{45}}^\rho) &&
y_1\sigma v_{\overline{5}}^{\phi}    
\end{array}\right) ~.
\eea
The matrix $M_u$ accommodates the three up quark masses and, combined
with $M_d$, leads to vanishing CKM mixing. The matrices $M_d$ and $M_e$ depend
only on four complex parameters, so that non-trivial relations are predicted among their
eigenvalues, as discussed after Eq.~(\ref{vd45}). If one insists to relax this constraint
in order to fit independently down quark and charged lepton masses, it suffices to
introduce an extra Higgs multiplet $\Delta' \sim (\overline{126},1,\alpha^m)$ with a VEV
in the $\overline{45}^{\Delta'}$ component. 
The neutrino mass matrices are given by
\beq
M_\nu = \frac{1}{\Lambda}\left( \begin{array}{ccccc} 
y_1\sigma   & &
y_3\chi   & & 
y_3\chi    \\ 
y_3\chi    & &
y_1\sigma   & &
y_3\chi    \\
y_3\chi    & & 
y_3\chi    & &
y_1\sigma     
\end{array}\right) v_5^{\phi} ~,\quad 
M_L = \frac{1}{\Lambda} \left( \begin{array}{ccc} 
f_1\tau  & 0 & f_3 \varphi  \\ 0 & f_1 \tau  & 0 \\  f_3  \varphi & 0 & f_1 \tau
\end{array}\right) v_{15}^{\Delta} ~,\quad 
M_R = \frac{1}{\Lambda} \left( \begin{array}{ccc} 
f_1\tau  & 0 & f_3 \varphi  \\ 0 & f_1 \tau  & 0 \\  f_3  \varphi & 0 & f_1 \tau
\end{array}\right) v_1^{\Delta} ~. 
\eeq
The light neutrino mass matrix has the form in 
Eq.~(\ref{type12}).
The departure from the TBM value of the $1-2$ lepton mixing angle is
controlled by the ratio
$
(v_5^\phi)^2/(v_1^\Delta v_{15}^\Delta) 
$,
where we assume that the couplings in 
Eq.~(\ref{supren}) and also the flavon VEVs in Eq.~(\ref{flavev}) 
are all of the same order. As discussed after Eq.~(\ref{s12}),
this ratio must be smaller than few percents, which is realized e.g.
for $v_5^\phi \approx 100$ GeV,
$v_1^\Delta \approx 10^{16}$ GeV and $v_{15}^\Delta \approx 0.1$ eV.

In the above model there is no mixing in the quark sector.
Let us briefly review possible mechanisms to generate
non-vanishing CKM parameters 
in $A_4$ models where $V_{CKM} = {\mathbbm 1}_3$ at leading order. 
In order to introduce the quark mixing, people considered 
(i) small explicit $A_4$ breaking Yukawas \cite{Ma:2002yp};
(ii) $A_4$ breaking in the soft supersymmetry breaking terms \cite {Babu:2002dz};
(iii) one-loop corrections that communicate to the quark sector the effect of
the $A_4$ spontaneous breaking in the $(0,1,0)$ direction 
\cite{He:2006dk,Bazzocchi:2007au}; 
(iv) spontaneous $A_4$ breaking
by a triplet with VEV  $(v,v,v_3)$ with $v\ne v_3$ \cite{Lavoura:2007dw}; 
(v) extra flavons transforming as $1'$ or $1''$ coupled to the quarks  
\cite{Altarelli:2008bg,Bazzocchi:2007na}.
In most of these cases 
it is problematic to generate  a sufficiently large quark mixing,
without introducing too large deviations from TBM lepton mixing \cite{Altarelli:2005yx}.
One possible exception is provided by  $SU(5)$ GUT models, 
where the CKM parameters can be introduced 
as a left-handed rotation in the down quark sector, which corresponds to 
a right-handed one in the charged lepton sector and therefore
does not affect lepton mixing \cite{Altarelli:2008bg}.
However, in our analysis of $SO(10)$ GUT models we did not find 
a contribution to $M_d$ and $M^T_e$ that introduces only a mixing on the left.
At the end of section \ref{sec:nonr} we will present a different 
mechanism to accommodate non-zero CKM
parameters, which does not perturb the TBM lepton mixing.

\section{Models with non-renormalizable operators \label{sec:nonr}}

In the previous section we have shown that
the TBM lepton mixing cannot be exactly realized 
by Yukawa couplings symmetric under $SO(10)\times A_4$.
In this section we solve this difficulty by considering the effect of
non-renormalizable operators that  contribute to the fermion mass matrices.
In addition, the analysis of such operators will provide a new tool
to accommodate the CKM parameters.

Higher dimensional operators are proportional to powers of $M_{GUT}/\Lambda$,
where the cutoff $\Lambda$ can be identified with the Planck or the string scale,
or with the mass of vector-like matter multiplets.
For definiteness, in the following we study this last possibility, by considering the
superpotential
\beq
W = y_{A} \Psi \Sigma \Phi_A + y_B \Psi \overline{\Sigma} \Phi_B 
- M \Sigma \overline{\Sigma} ~.
\label{sup5}\eeq
where $\Phi_A$ and $\Phi_B$ are Higgs multiplets whose components may acquire VEVs $a$ and $b$, respectively, $(\Sigma,\overline{\Sigma})$ is a vector-like
pair of matter multiplets and we assume $a,b \ll M$. In this case 
$\Sigma$ and $\overline{\Sigma}$ can be integrated out and one is left with
an effective operator
\beq
W^{eff} = \frac{y_Ay_B}{M}(\Psi \Phi_A)_\Sigma (\Psi \Phi_B)_{\overline{\Sigma}} ~,
\label{dim5}\eeq
where the subscripts specify how $\Psi$ and $\Phi_{A,B}$ are contracted.  
The VEVs $a$ and $b$ can be either of the order of $M_{GUT}$, if they
participate to the $SO(10)$ symmetry breaking to the SM, or
of the order of the electroweak scale, if they break the SM gauge group.

The flavour structure generated by the operator in Eq.~(\ref{dim5}) 
is determined by the $A_4$ assignments of $\Phi_A$, $\Phi_B$,
$\Sigma$ and $\overline{\Sigma}$. 
All the possibilities are analyzed in Appendix \ref{A2} and displayed in 
Table \ref{tab:fl5D}.
Beside the structures 
already possible with renormalizable Yukawa couplings, 
several new flavour structures can be realized
 (compare with Table \ref{tab:fl4D}).
This provides new options for model-building with $A_4$ symmetry.
Here we will focus on the realization of TBM lepton mixing, 
but alternative scenarios can be studied on the same footing.

By surveying all the possible operators listed in Table \ref{tab:fl5D},
one can identify the flavour structures compatible with the TBM mixing scheme. 
The cross structure of the neutrino Dirac and Majorana mass matrices,
shown in Eqs.~(\ref{mass}) and (\ref{seesawI}), can be built with the following components:
\bei
\item  $diag(1,1,1)$ from the operator A;
\item  
$\left(
\begin{array}{ccc}
0&0&1\\
0&0&0\\
1&0&0\\
\end{array}
\right)$ from the operator B with VEV alignment $(0,a,0)$
and no antisymmetric term ($y_A^a=0$);
\item  $diag(0,1,0)$ from the operator C (or C$'$) with VEV alignment
$(0,a,0)$ and $(0,b,0)$;
\item  
$\left(
\begin{array}{ccc}
1&0&1\\
0&0&0\\
1&0&1\\
\end{array}
\right)$ from the operator C with VEV alignment $(a,0,a)$ and $(b,0,b)$;
\item $diag(1,0,1)$ from the operator D with VEV alignment
$(0,a,0)$ and $(0,b,0)$ and either $y_{A,B}^a=0$ or $y_{A,B}^s=0$;
\item $\left(
\begin{array}{ccc}
1&0&\pm1\\
0&2&0\\
\pm1&0&1\\
\end{array}
\right)$ from the operator D with VEV alignment $(a,0,a)$ and $(b,0,b)$
and either $y_{A,B}^a=0$ or $y_{A,B}^s=0$.
\eei
The magic structure of the charged fermion mass matrices, shown in
Eq.~(\ref{mass}), can be built with the following components:
\bei
\item $diag(1,1,1)$ from the operator A;
\item  
$y_A^s\left(
\begin{array}{ccc}
0&1&1\\
1&0&1\\
1&1&0\\
\end{array}
\right) +  
y_A^a\left(
\begin{array}{ccc}
0&1&-1\\
-1&0&1\\
1&-1&0\\
\end{array}
\right)$
from the operator B with VEV alignment $(a,a,a)$;
\item 
$\left(
\ba{ccc}
1&1&1\\
1&1&1\\
1&1&1\\
\ea
\right)$ 
from the operator C with VEV alignment $(a,a,a)$ and $(b,b,b)$;
\item  
$\left(
\ba{ccc}
1&\omega^2&\omega \\
\omega&1&\omega^2 \\
\omega^2&\omega&1 \\
\ea
\right)$ 
from the operator C$'$ with VEV alignment $(a,a,a)$ and $(b,b,b)$;
\item 
$y_A^sy_B^s\left(
\ba{ccc}
2&1&1\\
1&2&1\\
1&1&2\\
\ea
\right) +  
y_A^ay_B^a\left(
\ba{ccc}
2&-1&-1\\
-1&2&-1\\
-1&-1&2\\
\ea
\right) +  
(y_A^sy_B^a-y_A^ay_B^s)\left(
\ba{ccc}
0&1&-1\\
-1&0&1\\
1&-1&0\\
\ea
\right)$
from the operator D with VEV alignment $(a,a,a)$ and $(b,b,b)$.
\end{itemize}

The contribution of the operator in Eq.~(\ref{dim5}) to the mass matrices of
the different sectors 
is determined by the $SO(10)$ assignments of $\Phi_A$, $\Phi_B$,
$\Sigma$ and $\overline{\Sigma}$. 
Higher dimensional operators have been often used to build realistic $SO(10)$ models without large representations, which may be disfavoured theoretically.
We thus perform a systematic analysis of dim-5 operators involving only $SO(10)$ multiplets of size less than or equal to 120.
There are only six such operators, that are analyzed in Appendix \ref{A2}
and displayed in Table \ref{tab:ga5D}.
Several qualitative new relations between the various sectors are possible,
with respect to the case of renormalizable Yukawa couplings
(compare with Table \ref{tab:ga4D}):
\bei
\item the operator $16_M16_M 16_H 16_H$  ($16_M16_M\overline{16}_H
\overline{16}_H$)
contributes only to the down (up) sector; here the subscripts distinguish
matter ($M$) from Higgs ($H$) multiplets;
\item the operator $16_M16_M\overline{16}_H\overline{16}_H$ can generate
Majorana masses for left-handed and right-handed neutrinos, with no need of
any $\overline{126}_H$ multiplet;
\item  
the operator $(16_M\overline{16}_H)_1(16_M\overline{16}_H)_1$
contributes to $M_\nu$ without affecting $M_u$;
\item the operator $(16_M 120_H)_{16} (16_M 45_H)_{\overline{16}}$ provides
two independent contributions to $M_\nu$ and  $M_u$, proportional to 
$\langle 5_H^{120}\rangle$ and $\langle 45_H^{120}\rangle$, respectively. 
\eei

We are now in the position to build a minimal  model
which realizes TBM lepton mixing using the dim-5 operators.
Notice first that, in the absence of $\overline{126}_H$ multiplets, 
the type I seesaw is the dominant source of light neutrino masses.
In fact, the type II contribution to $m_\nu$ generated by
the operator  $16_M16_M\overline{16}_H \overline{16}_H$
is of the order $\langle 5_H^{\overline{16}} \rangle^2 / M \lesssim 
(100\,{\rm GeV})^2/M_{GUT} \sim 10^{-3}$ eV, that is negligible.

In section \ref{sec:ren} we showed that, in the case of type I seesaw,
one cannot reproduce TBM lepton mixing by using renormalizable Yukawa couplings.
The technical reason is the absence of an operator that provides an
off-diagonal symmetric term, needed for the magic structure of $M_u$,
without modifying the cross structure of $M_\nu$.
This difficulty can be overcome by using  dim-5 operators.
The mechanism to generate the required contribution to $M_u$
is most easily described in $SU(5)$ language:
the VEV of the up-type doublet in a 45 Higgs multiplet couples 
(antisymmetrically) to two 10 matter multiplets generating $M_u$,
while it does not contribute to $M_\nu$.
In order to make this contribution to $M_u$ not antisymmetric,
one needs the insertion of the VEV of 
a 24 Higgs multiplet, which couples to 10 and $\overline{10}$ matter multiplets.
Since the SM singlet in 24 is in the hypercharge direction, 
the Clebsch-Gordan coefficients for the $Q$ and $u ^c$ components are different 
(by a relative factor $-4$), thus
making $M_u$ not antisymmetric.
This mechanism is embedded in $SO(10)$ 
by the operator
$(16_M 120_H)_{16} (16_M 45_H)_{\overline{16}}$ , with $45_H^{120}$ 
and $24_H^{45}$ acquiring a VEV.
By inspecting Table \ref{tab:ga5D}, this is in fact the only possible contribution to $M_u$
that does not affect $M_\nu$, at least with multiplets of dim $\le 120$.
Notice that such contribution to $M_u$ is symmetric when the VEV of $45_H$ is in the
$B-L$ direction (to see this, replace Eq.~(\ref{PS}) in the row VIII of 
Table \ref{tab:ga5D}).

For completeness, let us mention another $SU(5)$ mechanism 
that allows to generate only $M_u$ and not $M_\nu$:
one may employ the VEV of a 75 Higgs multiplet, which couples to
10 and  $\overline{10}$ (but not to 5 and $\overline{5}$) matter multiplets.
The minimal $SO(10)$ embedding is provided by the operator
$(16_M 10_H)_{16} (16_M 210_H)_{\overline{16}}$, with $5_H^{10}$ and
$75^{210}_H$ acquiring a VEV. 
However, the 75 component that acquires the VEV couples with opposite
Clebsch-Gordan coefficients to the $Q$ and $u^c$ components.
As a consequence the contribution to $M_u$ of this operator is antisymmetric. A symmetric contribution requires an extra antisymmetric coupling,
which can be introduced by replacing $10_H$ with $120_H$.

Our strategy to build an explicit model is to start from the renormalizable model 
discussed at the end of section \ref{sec:ren}, which is defined by the
superpotential in Eq.~(\ref{supren}), and replace some of the couplings with
the appropriate higher dimensional operators.
The field content of the model is given in Table \ref{scalarcontent}, where we introduced
a $Z_N$ symmetry. 
The Yukawa superpotential invariant under $SO(10)\times A_4\times Z_N$ is given by 
\beq
W_Y = \frac{y_1 \sigma}{\Lambda}\Psi\Psi\phi  
+ \frac{g_3 \chi}{\Lambda}\Psi\Psi\rho
+ \frac{h_3 \chi}{\Lambda}\Psi \overline{\Sigma}_2 A
- M_2 \overline{\Sigma}_2 \Sigma_2 + h \Sigma_2 \Psi \rho' 
+\frac{f_1 \tau  + f_3 \varphi}{\Lambda}\Psi \overline{\Sigma}_1 \xi 
- M_1  \overline{\Sigma}_1\Sigma_1 + f \Sigma_1 \Psi \xi
~.
\label{supNren}\eeq
In order to forbid all other couplings, up to terms quadratic in the flavons,
the $Z_N$ charges in Table \ref{scalarcontent} must be carefully chosen.
One viable choice is given by $N=8$, with $m=1$, $r=2$ and $n=5$.
By taking $N$ large enough, it is possible to forbid unwanted couplings 
up to higher order in the flavons.
Integrating out the heavy messenger fields $(\Sigma_i,\overline{\Sigma}_i)$, for $i=1,2$,
the effective superpotential takes the form
\beq
\label{Wy-noren}
W_Y^{eff} = 
\frac{y_1 \sigma}{\Lambda}\Psi\Psi\phi 
+ \frac{g_3 \chi}{\Lambda}\Psi\Psi\rho
+ \frac{h_3h\chi} 
{\Lambda} \, \frac{(\Psi \rho' )_{\Sigma_2} 
( \Psi A )_{\overline{\Sigma}_2}}{M_2}
+ \frac{(f_1 \tau +f_3\varphi)f}{\Lambda} \, \frac{(\Psi \xi)_{\Sigma_1}
(\Psi \xi)_{\overline{\Sigma}_1}}{M_1}  \,.
\eeq
For brevity, in Eqs.~(\ref{supNren}) and (\ref{Wy-noren}) 
we indicated with $h_3$  ($f_3$)  two independent couplings
$h_{3s}$ and $h_{3a}$ ($f_{3s}$ and $f_{3a}$), which correspond to the two
possible contractions of $A_4$ indexes in the product $3\times 3\times 3$
(see case B in Table \ref{tab:fl5D}).
The first operator in $W_Y^{eff}$ 
contributes to the Dirac mass matrices of the four sectors,
the second and the third only to the charged fermion mass matrices 
(we assume $\langle 5^{\rho} \rangle = \langle 5^{\rho'} \rangle = 0$) and 
the fourth only  to the Dirac and Majorana neutrino mass matrices.
The flavons take the same VEVs as in Eq.~(\ref{flavev}).

\begin{table}[t]
\begin{center}
\begin{tabular}{|c|c|c|c|}
\hline
& matter fields & Higgs fields & flavons \\
\cline{2-4}
\begin{tabular}{c} \\ $SO(10)$ \\ $A_4$ \\ $Z_N$ \\ \end{tabular} 
&
\begin{tabular}{ccccc} 
$\Psi$ & $\Sigma_1$ & $\overline{\Sigma}_1$ & $\Sigma_2$ & $\overline{\Sigma}_2$ \\  \hline 16 & 1 & 1 & 16 & $\overline{16}$ \\ 
3 & 3 & 3 & 3 & 3 \\ 
1 & $\alpha^{-r}$ & $\alpha^r$ & $\alpha^{-r}$ &  $\alpha^r$ \\ 
\end{tabular}
&
\begin{tabular}{ccccc}
$\phi$ & $\rho$ & $\rho'$ & $A$ & $\xi$ \\ \hline
$10$  & $120$ & $120$ & $45$ & $\overline{16}$ \\
1 & 1 & 1 & 1 & 1 \\
$\alpha^m$ & $\alpha^n$ & $\alpha^r$ & $\alpha^{n-r}$ & $\alpha^r$ \\
\end{tabular}
&
\begin{tabular}{cccc}
$\sigma$ & 
$\chi$ & $\tau$ & $\varphi$ \\ \hline
1 & 1 & 1 & 1  \\
1 & 3 & 1 & 3 \\
$\alpha^{-m}$ & $\alpha^{-n}$ &  
$\alpha^{-2r}$ & $\alpha^{-2r}$ \\
\end{tabular} 
\\ \hline\end{tabular}
\end{center}
\caption{Chiral superfields of a minimal non-renormalizable model for TBM mixing.
Here $\alpha^N=1$ and $1\le m\ne n\ne r < N$.}
\label{scalarcontent}
\end{table}

The charged fermion mass matrices acquire the magic structure
\beq
M_f = A_f {\mathbbm 1}_3 + \frac{B_f+C_f}{2} \left(\ba{ccc}
0 & 1 & 1 \\
1 & 0 & 1 \\
1 & 1 & 0 \\
\ea\right) + \frac{B_f-C_f}{2} \left
(\ba{ccc}
0 & 1 & -1 \\
-1 & 0 & 1 \\
1 & -1 & 0 \\
\ea\right) ~,
\eeq
where the coefficients have the following expression:
\bea
A_u &=&  \frac{y_1\sigma}{\Lambda}v_5^\phi ~,\nn\\
\frac{B_u+C_u}{2} &=& - \frac{h_{3s}h\chi}{\Lambda} 
\frac{5v_{24}^A}{M_2} v_{45}^{\rho'} ~,\nn\\
\frac{B_u-C_u}{2} &=&  \frac{g_3\chi}{\Lambda} v_{45}^\rho +
\frac{h_{3a}h\chi}{\Lambda} \frac{2v_1^A-3v_{24}^A}{M_2}
v_{45}^{\rho'} ~,\nn\\
A_d &=& \frac{y_1 \sigma}{\Lambda} v_{\overline{5}}^\phi ~,\nn\\
\frac{B_d+C_d}{2} &=& - \frac{h_{3s}h\chi}{\Lambda} 
\frac{4v_1^A-v_{24}^A}{M_2} (v_{\overline{5}}^{\rho'}+v_{\overline{45}}^{\rho'}) ~,\nn\\
\frac{B_d-C_d}{2} &=& \frac{g_3\chi}{\Lambda} 
(v_{\overline{5}}^\rho + v_{\overline{45}}^\rho)
- \frac{h_{3a}h\chi}{\Lambda} \frac{2v_1^A-3v_{24}^A}{M_2}
(v_{\overline{5}}^{\rho'}+v_{\overline{45}}^{\rho'}) ~,\nn\\
A_e &=& \frac{y_1 \sigma}{\Lambda} v_{\overline{5}}^\phi ~,\nn\\ 
\frac{B_e+C_e}{2} &=& - \frac{h_{3s}h\chi}{\Lambda} 
\frac{4v_1^A +9 v_{24}^A}{M_2} (v_{\overline{5}}^{\rho'}-3v_{\overline{45}}^{\rho'}) 
~,\nn\\
\frac{B_e-C_e}{2} &=&  -\frac{g_3\chi}{\Lambda}
(v_{\overline{5}}^\rho-3v_{\overline{45}}^\rho)
+  \frac{h_{3a}h\chi}{\Lambda} \frac{2v_1^A-3v_{24}^A}{M_2}
(v_{\overline{5}}^{\rho'}-3v_{\overline{45}}^{\rho'}) ~.
\label{array}\eea
These 9 quantities are independent, except for $A_e=A_d$, 
therefore the 9 masses of quarks and charged leptons can be accommodated
with this one constraint, that was already discussed in section \ref{sec:ren}.
The hierarchy between the three families requires 
$A_f$, $B_f+C_f$ and $B_f-C_f$ to be of the same order. In addition, since the
Yukawa of the top is close to one, in the up quark sector 
the three parameters should be  close to the electroweak scale. 
This implies that the flavon VEVs $\sigma$ and $\chi$ 
are not much smaller than $\Lambda$ 
and the VEV of the adjoint Higgs multiplet, $\vev{A}\sim M_{GUT}$, 
is not much smaller than $M_2$.
We will discuss in more detail the effect of vector-like matter fields
with mass close to $M_{GUT}$ in section \ref{sec:3h}.

The neutrino mass matrices take the form
\bea
\label{massmatr}
M_\nu &=& \frac{1}{\Lambda} \left( \begin{array}{ccc}  
y_1 \sigma v_5^\phi + 2 f_1 \tau  \frac{v_1^\xi}{M_1} v_5^\xi & 0 & 
2 f_{3s} \varphi \frac{v_1^\xi}{M_1} v_5^\xi \\ 0 & 
y_1 \sigma v_5^\phi + 2 f_1 \tau  \frac{v_1^\xi}{M_1} v_5^\xi & 0 \\ 
2 f_{3s} \varphi \frac{v_1^\xi}{M_1} v_5^\xi & 0 &  
y_1\sigma v_5^\phi +  2 f_1 \tau  \frac{v_1^\xi}{M_1} v_5^\xi 
\end{array}\right) ~,\nn\\ 
&&\nn\\
M_R &=& \frac{1}{\Lambda} \left(\begin{array}{ccc}
f_1\tau & 0 & f_{3s} \varphi \\ 
0 & f_1 \tau & 0  \\
f_{3s} \varphi & 0 & f_1\tau 
\end{array}\right) \frac{(v_1^\xi)^2}{M_1} ~,
\nn\\ 
M_L &=& \frac{1}{\Lambda} \left( \begin{array}{ccc}
f_1 \tau  & 0 & f_{3s}  \varphi \\
0 & f_1\tau & 0 \\
f_{3s}  \varphi & 0 & f_1 \tau 
\end{array}\right) \frac{(v_5^\xi)^2}{M_1} ~.
\eea
Light neutrino masses of the correct order of magnitude require 
$M_1\gg v_1^\xi\sim M_{GUT}$.
In this case the second term in Eq.~(\ref{seesaw}) can be sufficiently large
to accommodate oscillation data.
The TBM lepton mixing is exactly realized, as desired.

We conclude this section by showing that the dim-5 operators 
offer also the opportunity to generate non-zero quark mixing while preserving
TBM lepton mixing.
The idea is to introduce small deviations from the magic matrix structure in the
up quark sector. If these deviations were introduced in the down quark sector, that
would affect also the charged lepton mass matrix and, therefore, the TBM values of the
lepton mixing angles. We do not explore this possibility in this paper.

Interestingly, the Cabibbo  angle is naturally induced if the $A_4$ triplet VEV
aligned in the direction $(0,1,0)$, which is needed in the neutrino sector, 
also contributes to $M_u$.
This can be easily achieved by considering the same model as in Table
\ref{scalarcontent}, but with $\Sigma_1$ and $\overline{\Sigma}_1$ transforming 
as $45$ under $SO(10)$. In this case the fourth  operator in Eq.~(\ref{Wy-noren})
generates the neutrino sector as before (up to irrelevant Clebsch-Gordan factors,
see Table \ref{tab:ga5D}), but also contributes to the up quark sector
with
\begin{eqnarray}
\delta M_u&=& \frac{1}{\Lambda} \left( \begin{array}{ccc}   
f_1 \tau & 0 & f_{3s} \varphi \\ 
0 & f_1\tau & 0 \\
f_{3s} \varphi  & 0 & f_1 \tau  \end{array}\right)
  \frac{16 v_1^\xi v_5^\xi}{M_1} ~.
\label{deltaU}\end{eqnarray}
We absorb the diagonal entries of $\delta M_u$ in the definition of $A_u$ in
Eq.~(\ref{array}).  The off-diagonal entries, instead, do not respect the magic structure.
In the basis where $M_d$ is diagonal, the up quark mass matrix is given by
\begin{equation}
U^T_\omega (M_u^{magic} + \delta M_u) U^*_\omega =
\left( \begin{array}{ccc}   
m_{u1} + 2 \epsilon & -\omega^2 \epsilon & -\omega \epsilon \\ 
-\omega \epsilon & m_{u2} - \epsilon & 2\omega^2 \epsilon \\  
-\omega^2\epsilon & 2\omega \epsilon & m_{u3} -\epsilon 
\end{array}\right) ~,
\end{equation}
where $m_{ui}$ are defined in Eq.~(\ref{diag}) and
$\epsilon \equiv (16/3)(f_{3s}\varphi/\Lambda)(v_1^\xi/M_1) v_5^\xi$.
Notice that all entries of $M_u$ receive corrections of the same order.
Taking $\epsilon \ll m_{u2,3}$, which is natural since $v_1^\xi/M_1 \ll 1$,
one can accommodate the $1-2$ quark mixing angle, $\theta_{12}^q \approx 
|\epsilon/m_{u2}| \approx |\epsilon|/m_c$.
A similar idea to accommodate the Cabibbo angle was used in 
\cite{Bazzocchi:2007au}. 
The values of the other two quark mixing angles are given by
$\theta_{23}^q \sim \theta_{13}^q \sim (m_c/m_t) \theta_{12}^q$,
which unfortunately are two small to explain the experimental values.

It is possible to introduce other corrections to $M_u$ in such a way that all CKM
parameters can be accommodated, without affecting the structure of the other
mass matrices. For this purpose one needs a 120 Higgs multiplet $\rho_{up}$ 
with VEV only in the $SU(5)$ component $45$. Notice that in the model
of Table \ref{scalarcontent} one can make the identification $\rho\equiv \rho_{up}$,
since the VEVs $v_{\overline{5}}^\rho$ and $v_{\overline{45}}^\rho$ are not necessary
for the model to work.
Then, the operators
$\Psi\Psi\rho_{up}$ or $\Psi\Psi A\rho_{up}$ should couple to the flavon $\varphi$
in order to modify the magic structure of $M_u$. The flavour structure
of such corrections can be sufficiently rich to accommodate all
the CKM parameters, e.g. employing the operator D in Table \ref{tab:fl5D}.
However, we did not find any simple way to explain the hierarchy among the values
of the quark mixing angles, therefore we refrain from presenting further details.

\section{A mechanism to explain the inter-family mass hierarchy \label{sec:3h}}

In the previous sections, we analyzed the structure of fermion mixing.
We did not address yet the origin of the
strong hierarchy between the masses of the three families of quarks and charged
leptons.
In general, such hierarchy is not explained by the $SO(10)\times A_4$
symmetry by itself.
In fact, the mass matrix structure $M_f$ in Eq.~(\ref{mass}) accommodates 
three arbitrary mass eigenvalues, 
which are therefore free parameters like in the Standard Model.
Moreover, as we discussed in the Introduction,
since in $SO(10)$ both fermion chiralities transform
as 3 under $A_4$, one cannot use an extra family symmetry  besides $A_4$
in order to distinguish the three families.

The purpose of this section is to present an elegant mechanism to generate
the hierarchy between the three families in $SO(10)\times A_4$ models.
Such mechanism emerges from the analysis of dim-5 operators performed
in section \ref{sec:nonr} and it
turns out to be closely related to `universal seesaw' models
\cite{Bere83,Dimo83,DaWa,BeRa}.
We will show that it is compatible
with the generation of TBM lepton mixing discussed in the previous sections.

The masses of the three families of charged fermions are given in Eq.~(\ref{diag}).
They are linear combinations, with coefficients of unit modulus, of the mass matrix
elements $A_f,B_f,C_f$.
When the equality $A_f=\omega B_f = \omega^2 C_f \equiv m_{f3}/3$
holds, the mass eigenvalues are $(0,0,m_{f3})$.
When $A_f = (m_{f3} +  m_{f2})/3$, $B_f= (\omega^2 m_{f3} + \omega m_{f2})/3$ 
and $C_f= (\omega m_{f3}+\omega^2 m_{f2})/3$,
the mass eigenvalues are $(0,m_{f2},m_{f3})$.
If the above relations  were approximatively realized, one
might explain the inter-family hierarchy. At first sight such relations 
seem completely {\it ad hoc} because, in the $A_4$ models built so far,
the parameters $A_f,B_f,C_f$ are  generated by independent $A_4$ invariant
operators. To remove this problem, let us begin by writing the mass matrix structure
in Eq.~(\ref{mass}) as a sum over three rank-1 components:
\beq
M_f = \frac{m_{f1}}{3} \left(\ba{ccc}
1&1&1\\
1&1&1\\
1&1&1\\
\ea\right)
+ \frac{m_{f2}}{3} \left(\ba{ccc}
1&\omega&\omega^2\\
\omega^2&1&\omega\\
\omega&\omega^2&1\\
\ea\right)
+ \frac{m_{f3}}{3} \left(\ba{ccc}
1&\omega^2&\omega\\
\omega&1&\omega^2\\
\omega^2&\omega&1\\
\ea\right) ~.
\label{rank1}\eeq
The three flavour structures in Eq.~(\ref{rank1}) cannot be generated in
$A_4$ models with
renormalizable Yukawa couplings.  
However, we have shown in section \ref{sec:nonr} that several new
flavour structures are possible when the light matter fields $\Psi$  
mix with heavy matter multiplets. 
This mixing may actually lead to the three requisite structures
in Eq.~(\ref{rank1}).
Specifically, the ``democratic" structure is obtained 
from the dim-5 operator in Eq.~(\ref{dim5})
by taking $(\Sigma,\overline{\Sigma})$ singlets under $A_4$
and $\Phi_A$ and $\Phi_B$ transforming as
$A_4$ triplets with VEVs $(a,a,a)$ and $(b,b,b)$, 
corresponding to the case C in Table \ref{tab:fl5D}. 
The second (third) flavour structure in Eq.~(\ref{rank1})
is obtained similarly, but with the pair of matter multiplets 
$\Sigma'' \sim 1''$ and $\overline{\Sigma}'\sim1'$   
($\Sigma' \sim 1'$ and $\overline{\Sigma}''\sim1''$),
corresponding to the case C$'$ in Table \ref{tab:fl5D}.

There is a simple group theoretical interpretation of Eq.~(\ref{rank1}).
The $A_4$ representations decompose under the residual $Z_3$ symmetry
(preserved by the vacuum alignment $(1,1,1)$)
as $1_{A_4} = 1_{Z_3}$, $1'_{A_4}=1'_{Z_3}$, $1''_{A_4}=1''_{Z_3}$
and $3_{A_4} = 1_{Z_3}+1'_{Z_3}+1''_{Z_3}$.
The mass eigenstates are precisely those three orthogonal combinations 
of $\Psi_i \sim 3_{A_4}$ which  transform in a given representation
of $Z_3$ and, therefore, they separately mix  with $\Sigma$, $\Sigma'$ and 
$\Sigma''$, respectively. The three terms in  Eq.~(\ref{rank1}) correspond
to the $Z_3$ invariants $1_{Z_3}\times 1_{Z_3}$, $1'_{Z_3} \times 1''_{Z_3}$
and $1''_{Z_3} \times 1'_{Z_3}$.

For definiteness, we take the two Higgs VEVs 
$a\sim M_{EW}$ and $b\sim M_{GUT}$.
Since the first and second generation Yukawa couplings are much smaller than one,
they are well described by the dim-5 operators obtained decoupling
$(\Sigma,\overline{\Sigma})$ and  $(\Sigma'',\overline{\Sigma}')$
at the mass scales $M_1\gg M_2 \gg M_{GUT}$.
On the contrary, the third generation Yukawa couplings (in particular the top) are large,
therefore we are led to consider a vector-like pair of matter multiplets 
$(\Sigma',\overline{\Sigma}'')$ with mass $M_3$
of the same order as $M_{GUT}$.
In this case it is not appropriate to integrate out these states and treat their effect
in terms of higher dimensional operators.
We shall instead consider explicitly the mixing 
of $\Sigma'$ with $\Psi$  and
show how this generates the third term in Eq.~(\ref{rank1}).
The `inverse' hierarchy $M_1\gg M_2 \gg M_3$ may be justified e.g. by a
Froggatt-Nielsen $U(1)$ symmetry with different charges for 
$\Sigma$, $\Sigma''$ and $\Sigma'$.

The above discussion applies to any model with the specified
$A_4$ assignment of fields. In the case of $SO(10)$ models,
one should carefully choose the $SO(10)$ operators and symmetry breaking
pattern, in order to generate the structure of $M_{u,d,e}$ as in Eq.~(\ref{rank1})
and preserve, at the same time, the cross structure of  $m_\nu$.
Let us consider, to begin with, the matter multiplets $\Sigma' \sim (16,1')$ and
$\overline{\Sigma}'' \sim (\overline{16},1'')$, together with 
the Higgs multiplets $\phi_i \sim (10,3)$ and $A_i \sim (45,3)$
(the discussion below can be easily generalized to different $SO(10)\times A_4$
assignments of the fields). The superpotential reads
\beq
W = - M_3 \Sigma' \overline{\Sigma}'' 
+ y (\Psi_1\phi_1 + \omega \Psi_2\phi_2+\omega^2\Psi_3\phi_3) \Sigma' 
+ g (\Psi_1 A_1 + \omega^2 \Psi_2 A_2 +\omega\Psi_3 A_3) \overline{\Sigma}''
~. 
\label{supU}\eeq
The direct contribution to the light fermion mass matrices from 
the couplings $\Psi_i \Psi_j \phi_k$ is assumed to be sub-dominant.
It can be forbidden e.g. by a parity symmetry $Z_2$ under which $\Psi$, $\phi$ and
$A$ are odd.

We assume that $A_i$ acquire VEVs $(V,V,V)$, 
such that $\langle A_i \rangle \Psi_{i\alpha} = V k_\alpha \Psi_{i\alpha}$,
where $\alpha$ runs over the 16 components of $\Psi_i$ and
$k_\alpha$ are Clebsch-Gordan coefficients which depend on the 
$SO(10)$ direction associated with $V$.
Then, the following linear combination of the 16 multiplet components
becomes heavy:
\beq
\Psi_\alpha^h =  c_\alpha \Sigma'_\alpha - \frac{s_\alpha}{\sqrt{3}}
(\Psi_{1\alpha}+\omega^2\Psi_{2\alpha}+\omega\Psi_{3\alpha}) ~,
\eeq
where
\beq
c_\alpha\equiv \frac{M_3}{\sqrt{|M_3|^2+3|gVk_\alpha|^2}} ~,\qquad 
s_\alpha\equiv \frac{\sqrt{3} g V k_{\alpha}} {\sqrt{|M_3|^2+3|gVk_\alpha|^2}}~.
\eeq
Notice that $|c_\alpha|^2+|s_\alpha|^2=1$. 
The light fermions $\psi^l_{i\alpha}$, $i=1,2,3$, are given by the three linear combinations  orthogonal to $\Psi^h_\alpha$. 
The unitary $4\times 4$ mixing matrix is defined by
\beq
\left(\ba{c}\Psi_{1\alpha} \\ \Psi_{2\alpha} \\ \Psi_{3\alpha}\\ \hline 
\Sigma'_\alpha \ea\right)
= \left(\ba{ccc|c}
\quad &  & \quad  & -s^*_\alpha/\sqrt{3} \\
 & P_\alpha &  & -\omega s^*_\alpha/\sqrt{3} \\
 &  &  & -\omega^2 s^*_\alpha/\sqrt{3} \\ \hline
 & Q_\alpha &  & c_\alpha^*
\ea\right)
\left(\ba{c}\psi^l_{1\alpha} \\ \psi^l_{2\alpha}\\ \psi^l_{3\alpha}\\ \hline \Psi^h_\alpha \ea\right) ~.
\label{bigU}\eeq
Since the choice of basis for $\psi^l_{i\alpha}$ is arbitrary, the matrices $P_\alpha$
and $Q_\alpha$ are determined up to a $3\times 3$ unitary rotation from the right.
We choose the basis where $\psi^l_{i\alpha}\rightarrow \Psi_{i\alpha}$
in the limit $s_\alpha\rightarrow 0$. In this case one finds
\beq
P_\alpha =  {\mathbbm 1}_3+ \frac{|c_\alpha| -1}{3} \left(\ba{ccc}
1 & \omega^2 & \omega \\
\omega & 1 & \omega^2 \\
\omega^2 & \omega & 1 \\
\ea\right) ~, \qquad Q_\alpha = \frac{s_\alpha c^*_\alpha}{\sqrt{3}|c_\alpha|}
\left(\ba{ccc}1 & \omega^2 & \omega
\ea\right)~.
\eeq
In the following we take $c_{\alpha}$ real and positive, without loss of generality.

The three light states acquire a mass when the Higgs doublets $H^{u,d}_i$ in $\phi_i$
acquire VEVs at the electroweak scale. Let us begin from the down quark sector.
The second term  in Eq.~(\ref{supU}) contains the couplings 
$y (Q_1H_1^d+\omega Q_2 H_2^d + \omega^2 Q_3 H_3^d) d^c_{\Sigma'}$
as well as $y Q_{\Sigma'} (d^c_1 H^d_1+\omega d^c_2 H^d_2 + \omega^2 
d^c_3 H^d_3)$. 
Assuming the VEV alignment 
$(v_d,v_d,v_d)$ and using Eq.~(\ref{bigU}) for $\Psi_{i\alpha} = Q_i$ $(d_i^c)$
and $\Sigma'_\alpha = Q_{\Sigma'}$ ($d^c_{\Sigma'}$), we obtain
the following mass term for the three light generations:
\beq
W \supset 
(d_1^l~d_2^l~d_3^l)~ \frac{yv_d}{\sqrt{3}} \left[
s_Q \left(\ba{ccc}
1&\omega&\omega^2\\
\omega^2&1&\omega\\
\omega&\omega^2&1\\
\ea\right)
+\left(\ba{ccc}
1&\omega^2&\omega\\
\omega&1&\omega^2\\
\omega^2&\omega&1\\
\ea\right)
s_{d^c} 
\right] \left(\ba{c} d_1^{cl} \\ d_2^{cl} \\ d_3^{cl} \ea\right) ~.
\label{massd}\eeq
The charged lepton mass term has this same form with the obvious
replacements $Q\rightarrow L$ and $d^c \rightarrow e^c$.
The two flavour structures in Eq.~(\ref{massd}) are exactly those
required to reproduce the second and third generation masses
as in Eq.~(\ref{rank1}).
The relative hierarchy between the two terms is
determined by the direction of the $SO(10)$-breaking VEV of $A$.
In general $\langle A \rangle = v_3^A T_{3R} + v_{15}^A T_{B-L}$, where
$T_{3R}$ ($T_{B-L}$) is the generator of the
right-handed isospin (the $B-L$ symmetry). 
When $v_{15}^A=0$, the VEV of $A$ points in the right-handed isospin direction
and one has $s_Q=s_L=0$ as well as $s_{d^c}=s_{e^c}$. In this limit only the third
family is massive and $b-\tau$ unification is realized.

Let us now consider the up-quark and Dirac neutrino sectors. If also
$H^u_i$ acquire non-zero VEVs $(v_u,v_u,v_u)$,
then  $M_u$ ($M_\nu$) takes the same form as 
in Eq.~(\ref{massd}),
with the replacements $v_d\rightarrow v_u$ and $d^c \rightarrow u^c$ 
($Q\rightarrow L$, $d^c\rightarrow \nu^c$).
However, such contribution to $M_\nu$ 
would spoil the TBM lepton mixing.
The way to generate only $M_u$ and not $M_\nu$ is
the same described in section \ref{sec:nonr}: the up-type Higgs doublets which are kept light and acquire a VEV should not be those in $\phi\sim(10,3)$,
but rather those in the $SU(5)$ 45-components of $\rho \sim (120,3)$.
In this case the above derivation can be still 
applied replacing the coupling $y\Psi \Sigma' \phi$
by $h \Psi  \Sigma' \rho$.
When $v_{15}^A=0$,
only the top quark acquires a mass.

Next, let us discuss the masses of the second family of charged fermions.
One contribution may come from a non-zero $v_{15}^A \ll v_3^A$.
In this case $s_Q = -s_L/3 \sim (v_{15}^A/v_3^A) s_{d^c}$, so that 
the hierarchy between the two terms  in Eq.~(\ref{massd}) follows from a hierarchy
between the  $SU(4)_c$ and  the $SU(2)_R$ symmetry breaking scales.
(A recent Pati-Salam model to obtain a flavour hierarchy from
a gauge hierarchy without any family symmetry was recently proposed
in Ref.~\cite{FKR}.)
Notice also that the $s$ and $\mu$ masses are generated with the appropriate
Georgi-Jarlskog factor, due to their relative $B-L$ number.
However, if $v_{15}^A$ contributes also to the up quark sector,
one finds $m_c/m_t \sim m_s/m_b$, in strong disagreement with data.
The steepest hierarchy in the up quarks cannot be accommodated in this minimal
setup.

In fact, there is yet another reason to take the VEV of $A$ in the right-handed isospin
direction. We are working in the hypothesis that the cross structure of $m_\nu$
in Eq.~(\ref{mass}) is generated in the $\Psi_i$ basis.
In order to preserve the TBM mixing in the lepton sector,
such structure should not be changed by the rotation to the $\psi_i^l$ basis.
This is guaranteed by taking $v_{15}^A=0$,
because in this case $s_L=0$ and therefore the lepton doublets
$L_i$ do not mix with $L_{\Sigma'}$.

We set $v_{15}^A=0$ in the following. In this case, 
the second family masses may be introduced,
as already mentioned,
by the mixing with a pair $\Sigma'' \sim (16,1'')$
and $\overline{\Sigma}' \sim (\overline{16},1')$ with mass $M_2 \gg M_3$.
One may repeat the analysis done in Eqs.~(\ref{supU})-(\ref{massd})
with some obvious replacements, in particular swapping  
$\omega$ with $\omega^2$ everywhere.
Assuming that $H_d$ resides in $\phi$ and $H_u$ in $\rho$,
consistently with the generation of the third family masses, 
one can accommodate $m_c/m_t \ll m_s/m_b$ by tuning
the independent Yukawa couplings. Some more effort will be needed to make
$m_s\ne m_\mu$, as discussed later.

Analogously, one can generate the democratic term in Eq.~(\ref{rank1})
for the first family masses, by the mixing with a pair
$\Sigma \sim (16,1)$ and $\overline{\Sigma} \sim (\overline{16},1)$
with mass $M_1 \gg M_2$. The derivation is again very similar to
Eqs.~(\ref{supU})-(\ref{massd}),
replacing everywhere $\omega$ and $\omega^2$ with $1$.
Let us remind, however, that since the first family masses are tiny,
they may be generated even by non-democratic contributions,
from operators close to the Planck
scale, which would not affect significantly the values of the mixing angles.

Notice that no CKM mixing between the light quark families is generated by the
mechanism described above, as long as the $A_4$ VEV alignment $(1,1,1)$ is
preserved in all the operators contributing to $M_u$ and $M_d$.
In this case a $Z_3$ subgroup is unbroken and,
as mentioned above, each of the heavy matter multiplets $\Sigma,\Sigma',\Sigma''$ mixes only with the orthogonal combination of the $\Psi_i$ fields that transforms in the same way under $Z_3$. As a consequence, 
each light family of quarks does not mix with the others. 
In order to generate the CKM parameters, one should resort to extra contributions
to the quark mass matrices which break $Z_3$.

\begin{table}[t]
\begin{center}
\begin{tabular}{|c|c|c|c|}
\hline
& matter fields & Higgs fields & flavon \\
\cline{2-4}
\begin{tabular}{c} \\ $SO(10)$ \\ $A_4$ \\ $Z_N$ \\ \end{tabular} 
&
\begin{tabular}{ccccc} 
$\Psi$ & $\Sigma'$ & $\overline{\Sigma}''$ & $\Sigma''$ & $\overline{\Sigma}'$ 
\\ \hline 
16 & 16 & $\overline{16} $& 16 & $\overline{16}$  \\ 
3 & $1'$ & $1''$ & $1''$ & $1'$  \\ 
1 &  $\alpha^{-2m}$  & $\alpha^{2m}$ & $\alpha^{-2m}$ & $\alpha^{2m}$   \\ 
\end{tabular}
&
\begin{tabular}{ccc}
$\rho_1$& $\rho_2$ & $A$  \\ \hline
$120$ & $120$ & $45$  \\
1 & 1 & 1 \\
$\alpha^m$ & $\alpha^m$ & $\alpha^{-3m}$  \\
\end{tabular}
&
\begin{tabular}{c}
 $\chi$  \\ \hline
1 \\ 3  \\
$\alpha^{m}$ \\
\end{tabular} 
\\ \hline\end{tabular}
\end{center}
\caption{Chiral superfields needed 
to realize the inter-family mass hierarchy in the quark and charged lepton sectors.
Here $\alpha^N=1$ and $1\le m < N$.}
\label{scalarcontentV}
\end{table}

Let us implement in an explicit model the mechanism to generate the 
inter-family mass hierarchy together with TBM lepton mixing and non-zero 
Cabibbo mixing.
In order to correctly describe the masses of third and second family of quarks
and charged leptons, we found that the minimal set of multiplets 
is the one given in Table \ref{scalarcontentV}. 
The Yukawa superpotential has the form
\begin{eqnarray}
W_Y & = &  
\Psi \left(\frac{y_3\chi}{\Lambda}{\rho_1} +\frac{h_3\chi}{\Lambda} \rho_2 \right)
\Sigma'
- M_3 \Sigma' \overline{\Sigma}''
+  \overline{\Sigma}'' \frac{g_3\chi}{\Lambda}A \Psi   
\nn\\ 
& + & \Psi \left(\frac{y_2\chi}{\Lambda}{\rho_1} +\frac{h_2\chi}{\Lambda} \rho_2\right)\Sigma''
- M_2 \Sigma'' \overline{\Sigma}'
+  \overline{\Sigma}' \frac{g_2\chi}{\Lambda}A \Psi   ~. 
\label{supNrenV}
\end{eqnarray}
A $Z_N$ symmetry was introduced to forbid all other couplings, up to terms
quadratic in the flavons;
the minimal choice is $N=4$ with charge $m=1$.
The adjoint Higgs  multiplet $A$ has the VEV $v_3^A$ in the right-handed isospin 
direction and the triplet flavon has the VEV $\vev{\chi_i} = (\chi,\chi,\chi)$.
In order to reproduce correctly all the mass ratios, we introduced two 120 multiplets
$\rho_1$ and $\rho_2$ with the VEVs 
\beq\ba{l}
\vev{45^{\rho_1}} = v_{45}^{\rho_1} ~,~~~
\langle \overline{5}^{\rho_1} \rangle =v_{\overline{5}}^{\rho_1} ~,~~~
\vev{\overline{45}^{\rho_2}}=v_{\overline{45}}^{\rho_2} ~.
\\ 
\ea\eeq
Following the derivation of
Eqs.~(\ref{supU})-(\ref{massd}),
it is straightforward to compute the mass eigenvalues defined by Eq.~(\ref{rank1}):
\beq\ba{ll}
m_t = \displaystyle{\frac{\sqrt{3} y_3\chi}{\Lambda}} v_{45}^{\rho_1} s_3 ~,~~~~~ &
m_c = \displaystyle{\frac{\sqrt{3} y_2\chi}{\Lambda}} v_{45}^{\rho_1} s_2 ~,~~~ 
\\
m_b = - \left(\displaystyle{\frac{\sqrt{3} y_3\chi}{\Lambda}} v_{\overline{5}}^{\rho_1}
+ \displaystyle{\frac{\sqrt{3} h_3\chi}{\Lambda}} v_{\overline{45}}^{\rho_2} \right) s_3 
~,~~~~~ &
m_s = - \left(\displaystyle{\frac{\sqrt{3} y_2\chi}{\Lambda}} v_{\overline{5}}^{\rho_1}
+ \displaystyle{\frac{\sqrt{3} h_2\chi}{\Lambda}} v_{\overline{45}}^{\rho_2} \right) s_2 ~,~~~
\\
m_\tau = \left(\displaystyle{\frac{\sqrt{3} y_3\chi}{\Lambda}} v_{\overline{5}}^{\rho_1}
-3 \displaystyle{\frac{\sqrt{3} h_3\chi}{\Lambda}} v_{\overline{45}}^{\rho_2} \right) 
s_3 ~,~~~~~ &
m_\mu = \left(\displaystyle{\frac{\sqrt{3} y_2\chi}{\Lambda}} v_{\overline{5}}^{\rho_1}
-3 \displaystyle{\frac{\sqrt{3} h_2\chi}{\Lambda}} v_{\overline{45}}^{\rho_2} \right) s_2 ~.
\ea
\label{mass6}\eeq
where the parameters $s_3$, $s_2$ control the mixing between $\Psi$
and $\Sigma'$, $\Sigma''$ and are given by
\beq
s_3 \equiv \frac{\sqrt{3}g_3\chi v_3^A}{\sqrt{|M_3\Lambda|^2+
| \sqrt{3}g_3\chi v_3^A |^2}} ~,~~~~~
s_2 \equiv \frac{\sqrt{3}g_2\chi v_3^A}{\sqrt{|M_2\Lambda|^2+
| \sqrt{3}g_2\chi v_3^A |^2}} ~.
\eeq
The heaviness of the top requires to take $M_3\sim v_3^A$ so that
$s_3 \sim 1$. The hierarchy between second and third generation masses is then
explained by taking $M_2 \gg v_3^A$, so that $s_2\ll 1$.
Approximate $b-\tau$ unification is realized when the first term in $m_b$ and $m_\tau$
dominates over the second.
The ratio $m_s/m_\mu\sim 1/3$ is realized when the second term in $m_s$ and
$m_\mu$ dominates over the first.
More precisely, the six masses in Eq.~(\ref{mass6}) are constrained by one non-trivial
relation, $(3m_s - m_\mu) = (m_c/m_t)(3m_b-m_\tau)$, which connects the two
phenomenological facts $m_s/m_b\gg m_c/m_t$  and  $m_\mu\approx 3m_s$ at the
GUT scale.
At this level the first family masses are vanishing.

As for the neutrino sector, we need to generate the cross structure of $m_\nu$,
as required by TBM mixing. 
One may think that is sufficient to introduce
the last operator in Eq.~(\ref{Wy-noren}),
however the size of neutrino masses is too small in this case, as can be seen 
by inspecting  Eq.~(\ref{massmatr}) (in the present scenario 
the term $y_1\sigma v_5^{\phi}$ in $M_\nu$ is absent).
To solve this problem,
the neutrino Dirac mass matrix
$M_\nu$ and the right-handed neutrino Majorana mass matrix $M_R$
should be generated by two independent operators.
We found that the minimal set of multiplets to achieve this purpose
is the one given in Table \ref{scalarcontentVbis}, 
with two $\overline{16}$ Higgs multiplets taking VEVs only in the
directions $\vev{1^{\xi_M}} = v_1^{\xi_M}$ and $\vev{5^{\xi_D}} = v_5^{\xi_D}$.
We introduced an auxiliary symmetry 
$Z_{N'}$ such that all fields charged under $Z_N$ are neutral under
$Z_{N'}$ and vice versa.
Then, the superpotential in Eq.~(\ref{supNrenV}) is extended to include
the extra terms 
\begin{eqnarray}
W_Y & \supset & \Psi \left( \frac{f_{1M} \tau  + f_{3M} \varphi}{\Lambda} \xi_M 
+ \frac{f'_{\sigma} \sigma}{\Lambda}\xi_D \right) \overline{\Sigma}_M 
- M_M  \overline{\Sigma}_M \Sigma_M + f_M \Sigma_M  \xi_M \Psi
\nn\\
&+& 
\Psi\, \frac{f_{\sigma}\sigma}{\Lambda} \xi_M  \overline{\Sigma}_D 
- M_D  \Sigma_D \overline{\Sigma}_D 
+ f_D  \Sigma_D \xi_D  \Psi
~. \label{supNrenVbis}
\end{eqnarray}
The $Z_{N'}$ assignments of Table \ref{scalarcontentVbis}
forbid all other $SO(10)\times A_4$ invariant couplings
(a minimal viable choice is $N'=5$, with charges $n=2$ and $r=4$).
Once the messenger fields $(\Sigma_D,\overline{\Sigma}_D)$ and 
$(\Sigma_M,\overline{\Sigma}_M)$
are integrated out, the effective superpotential contains two relevant operators:
\beq
W_Y^{eff} \supset
\frac{(f_{1M} \tau  + f_{3M} \varphi)f_M}{\Lambda}   
\frac{(\Psi \xi_M)_{\overline{\Sigma}_M}(\Psi \xi_M)_{\Sigma_M}}{M_M}
+ \frac{f_\sigma f_D \sigma}{\Lambda}   
\frac{(\Psi \xi_M)_{\overline{\Sigma}_D}
(\Psi \xi_D)_{\Sigma_D}}{M_D}
~.
\label{nueff}\eeq
The first operator generates the cross structure of $M_R$,
while the second  generates $M_\nu \propto \mathbbm{1}_3$ (remember that only the
singlet in $\xi_M$ and the doublet in $\xi_D$ acquire a non-zero VEV).
By taking $M_M \gg M_D \ge M_{GUT}$, 
one obtains right-handed neutrino masses significantly below $M_{GUT}$
and keeps the Dirac neutrino masses close to electroweak scale. 
In this way sufficiently large neutrino masses may be generated by type I seesaw.

\begin{table}[t]
\begin{center}
\begin{tabular}{|c|c|c|c|}
\hline
& matter fields & Higgs fields & flavons \\
\cline{2-4}
\begin{tabular}{c} \\ $SO(10)$ \\ $A_4$ \\ $Z_{N'}$ \\ \end{tabular} 
&
\begin{tabular}{ccccc} 
$\Psi$ & $\Sigma_D$ &$\overline{\Sigma}_D$ & $\overline{\Sigma}_M$ & $\Sigma_M$ 
\\ \hline 
16 & 1 & 1 & 45 & 45 \\ 
3   & 3 & 3 & 3  & 3 \\ 
1  & $\alpha^{-n}$ & $\alpha^n$ & $\alpha^{-r}$ & $\alpha^r$  \\ 
\end{tabular}
&
\begin{tabular}{ccc}
$\xi_D$& $\xi_M$ \\ \hline
$\overline{16}$ & $\overline{16}$ \\
1 & 1 \\
$\alpha^n$ & $\alpha^{-r}$ \\
\end{tabular}
&
\begin{tabular}{ccc}
$\sigma$ & $\tau$ & $\varphi$ \\ \hline
1 & 1 & 1 \\
1 & 1 & 3 \\
$\alpha^{r-n}$ & $\alpha^{2r}$ & $\alpha^{2r}$ \\
\end{tabular} 
\\ \hline\end{tabular}
\end{center}
\caption{Chiral superfields needed to generate the cross structure of the neutrino
mass matrix and the Cabibbo mixing angle.
Here $\alpha^{N'}=1$ and $1\le n\ne  r < N'$.}
\label{scalarcontentVbis}
\end{table}

When also $\vev{5^{\xi_M}}=v_5^{\xi_M}$ is different from zero,
the first operator in Eq.~(\ref{nueff}) gives a negligible contribution to $M_\nu$,
because $M_M\gg M_D$. However, it plays an important role in the up
quark sector, since it provides a small correction to $M_u$ in a form 
completely analog to Eq.~(\ref{deltaU}). This correction generates the Cabibbo
mixing angle, in the same way as discussed at the end of section \ref{sec:nonr}.

In summary, the  model defined by the superpotential in Eqs.~(\ref{supNrenV})
and (\ref{supNrenVbis}) accounts for 
TBM mixing in the lepton sector and the $1-2$ mixing in the quark sector,
as well as for the hierarchical values of the quark and charged
lepton masses.

Before concluding, let us remark that,
in most models of `universal seesaw' with a realistic phenomenology,
the mass terms  $M \Sigma\overline{\Sigma}$ of the messenger fields
are sensitive to the $SO(10)$ breaking VEVs, thus giving different masses to the different
components of $\Sigma$. One may also introduce a non-trivial mixing between
the heavy matter families.
In our construction we barred these extra possibilities for simplicity.
A recent model of `universal seesaw' 
making use of rank-1 flavour structures, together with more references, can be found in \cite{BeNe}. Heavy messenger fields are also employed in a similar fashion
in an $SO(10)\times SU(3)$ model for TBM mixing \cite{deMedeirosVarzielas:2005ax}.

\section{Conclusions \label{sec:con}}

In this paper we performed a systematic analysis of fermion mass matrices
in models of Grand Unification based on the gauge symmetry $SO(10)$
with the discrete family symmetry $A_4$. In these models all light fermions and
right-handed neutrinos may be unified in a single multiplet $(16,3)$
of the group $SO(10)\times A_4$. We demonstrated that,
even though this scenario is very constrained,
it is possible to understand the disparity between
the quark and lepton mixing angles as well as the strong
hierarchy between the three families of charge fermion masses.

In models with renormalizable Yukawa couplings,
we found that the exact TBM lepton mixing
can be obtained only if the type I seesaw contribution to $m_\nu$ is neglected.
The specific effect of such contribution
is to shift the $1-2$ lepton mixing angle from the
tri-maximal value. The non-zero values of the CKM mixing angles cannot be
accommodated without introducing further departures from TBM lepton mixing.

A much richer flavour structure appears once the effect of higher dimensional
operators is taken into account. We considered dim-5 operators generated by the
mixing of light families with heavy vector-like matter multiplets:
the corresponding structures of the mass matrices are listed in Appendix \ref{A2},
for all $A_4$ representations and all the $SO(10)$ representations of size
$\le120$. This classification proves to be a useful tool for model-building.

We found that the dim-5 operators help to evade several difficulties
in the construction of a satisfactory $SO(10)\times A_4$ model of flavour.
One crucial ingredient is provided by those operators which contribute differently
to the Dirac mass matrices $M_u$ and $M_\nu$.
First, it is possible to obtain exact TBM lepton mixing from a type I seesaw.
Second, there are contributions  which generate non-zero CKM mixing angles
without disturbing the TBM pattern in the lepton sector.

Moreover, we have shown that the mixing of the three light families
with vector-like matter multiplets provides a natural explanation of the 
inter-family mass hierarchy of quarks and charged leptons.
In fact, rank-1 contributions to the mass matrices 
are generated by the mixing with heavy dim-16 multiplets,
that transform in a dim-1 representation of $A_4$.
The flavour structures
of such contributions are exactly those necessary to achieve TBM mixing.
A hierarchy in the masses of the heavy families and/or in the VEVs breaking
$SO(10)$ reflects into a hierarchy of the masses of the light families.

\section*{Acknowledgments}

We thank Jos\'e F. W. Valle for useful comments and encouragement.
MF thanks the Aspen Center for Physics
(2007 Workshop on ``Neutrino physics: looking forward"),
where this project was conceived by discussing with Ernest Ma and Jos\'e F. W. 
Valle, and the Instituto de F\'isica Corpuscular of Valencia, for the kind hospitality.
The work of FB and SM was supported by 
MEC grant FPA2005-01269 and
FPA2008-01935-E, by EC RTN network MRTN-CT-2004-503369,
and by Generalitat Valenciana ACOMP06/154. 
MF was supported in part by the Marie Curie
Intra-European Fellowship MEIF-CT-2007-039968, the CNRS/USA
exchange grant 3503, and
the RTN European Program MRTN-CT-2004-503369.

\appendix
\section{The group $A_4$ and its breaking pattern \label{B1}}

The pair of lepton mass matrix structures defined in in Eq.~(\ref{mass})
leads to exact  TBM mixing in the lepton sector.
In this Appendix we prove that such realization of TBM mixing is the only one
that can be obtained by the spontaneous breaking of an $A_4$ flavour symmetry.
We follow an approach already applied to the group $A_4$ in Ref.~\cite{He:2006dk} 
and to the group $T'$ in Ref.~\cite{Feruglio:2007uu} (see also \cite{Lam}).

The discrete group $A_4$ is formed by the even permutations of four objects. 
It can be defined by  two generators  $S$ and $T$ such that
\begin{equation}
\label{rel}
 S^2= T^3= (ST)^3=1\,.
 \end{equation}
The 12 elements of $A_4$ belong to four conjugacy classes
\begin{eqnarray}
\mathcal{C}_1&:& I ~;\nn\\
\mathcal{C}_2&:& T,ST, TS, STS ~;\nn\\
\mathcal{C}_3&:& T^2,ST^2, T^2S, TST ~;\nn\\
\mathcal{C}_4&:& S, T^2ST,TST^2 ~.
\end{eqnarray}
Each element $g$ of $\mathcal{C}_{2,3}$ ($\mathcal{C}_4$)
satisfies $g^3=1$ ($g^2=1$)
and, therefore, generates a subgroup $Z_3$ ($Z_2$) of $A_4$. 
There are three dim-1 and one dim-3 irreducible representations.
We adopt the basis where the generators  of the dim-3 representation are given by  
\beq
S= \left(\begin{array}{ccc} -1&0&0\\ 0&1&0  \\ 0&0  &-1\end{array} \right) ~,\qquad
T=\left(\begin{array}{ccc} 0&1&0\\ 0&0&1  \\ 1&0  &0\end{array} \right)~.
\eeq

We assume that both lepton doublets and singlets transform as $A_4$ triplets,
as dictated by $SO(10)$.
The lepton mixing matrix $U_{lepton}= U^\dag_e U_\nu$ depends on how 
$A_4$ is spontaneously broken in the charged lepton and in the neutrino sector. 
If $A_4$ were unbroken in one sector, 
the mass matrix would be proportional to the identity.
If $A_4$ were broken completely in one sector, 
the mass matrix would be generic and the values
of the mixing angles arbitrary.
Therefore, the only non-trivial case occurs when $A_4$ is broken to 
a subgroup in the charged lepton sector and to another subgroup in the neutrino sector.
There are three possibilities:
\bei
\item the VEV of an $A_4$ triplet with alignment $(1,1,1)$ breaks $A_4$ to $Z_3$;
\item the VEV of an $A_4$ triplet with alignment $(0,1,0)$ breaks $A_4$ to $Z_2$;
\item the VEV of a singlet in the $1'$ or $1''$ representation breaks $A_4$ to
$Z_2\times Z_2$.
\eei
The relevant choice to obtain TBM mixing is to preserve
$Z_3$ in one sector and $Z_2$ in the other, as we now show
(one may check in a similar way that all other choices do not lead to TBM mixing).

If $g$ is the generator of a subgroup of  $A_4$, the mass term $\psi M \psi^c$ 
is invariant with respect  to $g$ if and only if $gMg^T= M$.
For definiteness, and without loss of generality, let us consider the $Z_3$ and $Z_2$
subgroups generated by $T$ and $S$, respectively.
There are  two possibilities
\bea
\label{struc}
(i) \quad T\, M_e \, T^T= M_e\quad \Rightarrow \quad M_e=  \left(\begin{array}{ccc}
A_e & B_e & C_e \\ C_e & A_e & B_e \\ B_e & C_e & A_e \end{array} \right); &&
S\, m_\nu \, S^T = m_\nu \quad \Rightarrow \quad m_\nu =  \left(\begin{array}{ccc} a&0&b\\ 0&c&0 \\ b&0 &d\end{array} \right);\nn\\
(ii) \quad S\, M_e \, S^T= M_e\quad \Rightarrow \quad M_e=  \left(\begin{array}{ccc}
A_e & 0 & C_e \\ 0 & D_e & 0 \\ B_e & 0 & E_e \end{array} \right); &&
T\, m_\nu \, T^T = m_\nu \quad \Rightarrow \quad m_\nu =  \left(\begin{array}{ccc} a&b&b\\ b&a&b \\ b&b &a\end{array} \right).
\eea
In the case $(i)$   we recognize the charged lepton mass matrix $M_e$
of Eq.~(\ref{mass}). 
Notice that $m_\nu$ is more general than in Eq.~(\ref{mass}), because the $Z_2$
invariance allows different entries on the diagonal, in particular $a\neq d$. 
This is because one may couple to the neutrinos also a Higgs
transforming as $1'$ or $1''$, and still maintain $m_\nu$ invariant under $Z_2$ 
\cite{He:2006dk}. 
Therefore 
this $A_4$ breaking pattern is not sufficient to predict TBM mixing. 
If one does not  introduce such $1'$ and $1''$ fields in the model, one finds 
$a=c=d$ and the neutrino mass matrix of Eq.~(\ref{mass}) is recovered in the special 
limit $a=c$. Then exact TBM lepton mixing is obtained.

For what concerns the case $(ii)$,  one can obtain TBM mixing
in the neutrino sector alone,
by taking $M_e$ diagonal.
This scenario is realized  by coupling to the charged leptons three Higgs singlets
$1,1',1''$  and no triplet (the $A_4$ subgroup $Z_2\times Z_2$ is preserved in this
case). However, even though $m_\nu$ is diagonalized by $U_{TBM}$,
one has $m_1=m_3$,
in strong disagreement with oscillation experiments.
Therefore the case (ii) is not viable.

\section{Structures of mass matrices with $SO(10)\times A_4$ symmetry}

\subsection{Renormalizable Yukawa couplings \label{A1}}

Consider the Yukawa couplings $Y_{ij} \Psi_i \Psi_j \Phi$, where the matter multiplets transform as  $\Psi \sim (16,3)$ under $SO(10) \times A_4$ and $\Phi$ is
a Higgs multiplet. 
For all possible
$SO(10)\times A_4$ assignments of $\Phi$,
let us analyze the structure of the quark and lepton mass matrices that are generated
when a given component of $\Phi$ acquires a VEV $v$.

The $A_4$ assignment of $\Phi$ determines the flavour structure of the mass matrices, as shown in Table \ref{tab:fl4D}. When $\Phi \sim 3$ (case B in the Table),
there are two $A_4$-invariants, one symmetric and the other antisymmetric in the
flavour indexes. 
The $SO(10)$ assignment of $\Phi$ determines the
relative contribution to the mass matrices of the different sectors, as shown in Table
\ref{tab:ga4D}.
The Yukawa coupling matrices $Y_{10}$ and $Y_{\overline{126}}$ ($Y_{120}$) are (anti)symmetric in the flavour indexes.

For a given $SO(10)\times A_4$ assignment of $\Phi$,
the contribution to the mass matrices is obtained combining
the corresponding rows in Tables \ref{tab:fl4D} and \ref{tab:ga4D}.
Such contribution is non-zero only if both rows contain terms
(anti)symmetric in the flavour indexes.

\begin{center}
\begin{table}[t]
\begin{tabular}{|c|c|c|}
\hline
case & $A_4$ operator & mass matrix structure \\
\hline \hline
A & $3_M 3_M 1_H$ & $ y~diag (1,1,1)~v $\\
\hline 
A$'$ & $3_M 3_M 1'_H$ & $ y'~diag (1,\omega,\omega^2)~v'$\\
\hline 
A$''$ & $3_M 3_M 1''_H$ & $ y''~diag (1,\omega^2,\omega)~v''$\\
\hline
B& $3_M 3_M 3_H$ & $ 
y^s \left(\begin{array}{ccc} 0& v_3& v_2\\
v_3& 0& v_1\\
v_2& v_1&0\end{array}  \right) +
y^a \left(\begin{array}{ccc} 0& v_3& -v_2\\
-v_3& 0& v_1\\
v_2& -v_1&0\end{array}  \right)
$\\
\hline
\end{tabular}
\caption{The possible $A_4$ flavour structures arising from renormalizable
Yukawa couplings. The subscript $M$ ($H$) identifies the matter (Higgs) multiplets.}
\label{tab:fl4D}
\end{table}
\end{center}

\begin{center}
\begin{table}[t]
\begin{tabular}{|c|c|c|}
\hline 
case & $SO(10)$ operator & mass matrices \\
\hline \hline
I & $ 16_M 16_M 10_H$ & 
$\begin{array}{l}
M_u=M_{\nu}= Y_{10} v_5 \\
M_d=M_e = Y_{10} v_{\overline{5}} 
\end{array}$\\
\hline
II & $16_M 16_M 120_H$ & 
$\begin{array}{l}
M_u= Y_{120} v_{45} \\
M_{\nu} = Y_{120} v_5 \\
M_d = Y_{120} (v_{\overline{5}}+v_{\overline{45}}) \\
M^T_e = Y_{120} (v_{\overline{5}}-3v_{\overline{45}})
\end{array}  $\\
\hline
III & $16_M 16_M \overline{126}_H$ & 
$ \begin{array}{l}
M_u= Y_{\overline{126}}\, v_5 \\ 
M_{\nu} = -3 Y_{\overline{126}}\, v_5 \\  
M_d = Y_{\overline{126}}\, v_{\overline{45}} \\ 
M_e = -3 Y_{\overline{126}}\, v_{\overline{45}} \\ 
M_L = Y_{\overline{126}}\, v_{15} \\
M_R = Y_{\overline{126}}\, v_1
\end{array}$\\
\hline
\end{tabular}
\caption{The contributions to the mass matrices from $SO(10)$ invariant 
renormalizable Yukawa couplings. Different VEVs of the same $SO(10)$ Higgs multiplet carry a subscript indicating the $SU(5)$ component they belong to.} 
\label{tab:ga4D}
\end{table}
\end{center}

\subsection{Non-renormalizable  operators \label{A2}}

Consider the generic dimension-5 operator 
$c_{ij} \Psi_i \Psi_j \Phi_A \Phi_B / M$, where $\Phi_{A,B}$
are two (possibly equal) Higgs multiplets and $M$ is a cutoff scale.
When the appropriate components of $\Phi_A$ and $\Phi_B$ 
acquire VEVs $a$ and $b$, respectively, 
one generates contributions to the quark and lepton mass matrices.
Such operator may arise from the exchange of a pair of vector-like matter multiplets 
$(\Sigma,\overline{\Sigma})$ of mass $M$, with the superpotential 
given in Eq.~(\ref{sup5}).
If one is interested in physics at scales much smaller than $M$ (that is, for $a,b\ll M$),
the pair 
$(\Sigma,\overline{\Sigma})$ can be integrated out and one is left with the effective
superpotential
\beq
W^{eff} = 
\frac{y_A y_B}{M} (\Psi\Phi_A)_\Sigma (\Psi\Phi_B)_{\overline{\Sigma}} ~.
\label{d5}\eeq
The dim-5 operators may also arise from the exchange of
heavy Higgs multiplets $\Phi'$. We do not consider this possibility,
since in this case the operators involve the Yukawa coupling
$Y_{ij}\Psi_i\Psi_j \Phi'$, therefore they do not generate new mass matrix structures,
besides those already discussed in Appendix \ref{A1}.

The $A_4$ assignments of $\Phi_{A,B}$ and $(\Sigma,\overline{\Sigma})$
determine the flavour structure of the mass matrices.
There are 9 possible cases listed in Table 
\ref{tab:fl5D}.
The 5 cases B$'$, B$''$, C, C$'$, D provide new flavour structures with respect to
the Yukawa couplings in Table \ref{tab:fl4D}. Let us 
illustrate some new possibilities:
\bei
\item In the case C (and C$'$) each entry of the mass matrix is proportional to a different pair of VEVs, therefore matrices with only one non-zero entry (row, column) can be generated choosing the appropriate VEV alignment.
\item In the case D, by taking the alignment $(0,a,0)$ and $(0,b,0)$ 
only the entries (11) and (33) are generated, with relative coefficients $(y_A^s \pm y_A^a)(y_B^s\pm y_B^a)$.
\item In the case D, by taking the alignment $(a,a,a)$ and $(b,0,0)$, only the entries
$12,~13,~22,~33$ are generated;
if $y_{A,B}^a=0$, the 4 entries are equal.
\eei

\begin{table}[hbtp]\begin{center}
\begin{tabular}{|c|c|c|c|c|}
\hline
case & $A_4$ operator & mass matrix structure
\\ \hline \hline
A & $\ba{c} (3_M 1_H)_3 (3_M 1_H)_3 \\
 (3_M 1'_H)_3 (3_M 1''_H)_3 \ea$  & $y_A y_B~diag(1,1,1) \displaystyle{\frac{ab}{M}}$ \\
\hline
A$'$ & $\ba{c} (3_M 1_H)_3 (3_M 1'_H)_3 \\
 (3_M 1''_H)_3 (3_M 1''_H)_3 \ea$  & $y'_A y'_B~diag(1,\omega,\omega^2) 
 \displaystyle{\frac{ab}{M}}$ 
\\
\hline
A$''$ & $\ba{c} (3_M 1_H)_3 (3_M 1''_H)_3 \\
 (3_M 1'_H)_3 (3_M 1'_H)_3 \ea$  & $y''_A y''_B~diag(1,\omega^2,\omega) 
 \displaystyle{\frac{ab}{M}}$
\\
\hline
B & $(3_M 3_H)_3 (3_M 1_H)_3$ &
$\left[ 
y_A^s y_B \left(\begin{array}{ccc} 0& a_3& a_2\\
a_3& 0& a_1\\
a_2& a_1&0\end{array}  \right) +
y_A^a y_B \left(\begin{array}{ccc} 0& a_3& -a_2\\
-a_3& 0& a_1\\
a_2& -a_1&0\end{array}  \right)\right]
\displaystyle{\frac{b}{M}}$ \\
\hline
B$'$ & $(3_M 3_H)_3 (3_M 1'_H)_3$ &
$\left[ 
y_A^s y_B \left(\begin{array}{ccc} 0& a_3\omega& a_2\omega^2\\
a_3& 0& a_1\omega^2 \\
a_2& a_1\omega&0\end{array}  \right) +
y_A^a y_B \left(\begin{array}{ccc} 0& a_3\omega& -a_2\omega^2\\
-a_3& 0& a_1\omega^2 \\
a_2& -a_1\omega&0\end{array}  \right)\right]
\displaystyle{\frac{b}{M}}$ \\
\hline
B$''$ & $(3_M 3_H)_3 (3_M 1''_H)_3$ &
$\left[ 
y_A^s y_B \left(\begin{array}{ccc} 0& a_3\omega^2& a_2\omega\\
a_3& 0& a_1\omega\\
a_2& a_1\omega^2&0\end{array}  \right) +
y_A^a y_B \left(\begin{array}{ccc} 0& a_3\omega^2& -a_2\omega\\
-a_3& 0& a_1\omega\\
a_2& -a_1\omega^2&0\end{array}  \right)\right]
\displaystyle{\frac{b}{M}}$ \\
\hline
C & $(3_M 3_H)_1 (3_M 3_H)_1$ &
$y_A y_B \left( \ba{ccc}
a_1b_1 & a_1b_2 & a_1b_3 \\
a_2b_1 & a_2b_2 & a_2b_3 \\
a_3b_1 & a_3b_2 & a_3b_3 
\ea \right) \displaystyle{\frac{1}{M}}$
\\ \hline
C$'$ & $(3_M 3_H)_{1'} (3_M 3_H)_{1''}$ &
$y_A y_B \left( \ba{ccc}
a_1b_1 & a_1b_2\omega^2 & a_1b_3\omega \\
a_2b_1\omega & a_2b_2 & a_2b_3\omega^2 \\
a_3b_1\omega^2 & a_3b_2\omega & a_3b_3 
\ea \right) \displaystyle{\frac{1}{M}}$
\\ \hline
D & $(3_M 3_H)_3 (3_M 3_H)_3$ &
$\ba{c}
\left[
y^s_A y^s_B \left( \ba{ccc}
a_2b_2+a_3b_3 & a_2b_1 & a_3b_1 \\
a_1b_2 & a_3b_3+a_1b_1 & a_3b_2 \\
a_1b_3 & a_2b_3 & a_1b_1+a_2b_2 
\ea \right)
\right. \\ \left.
+y^a_A y^a_B \left( \ba{ccc}
a_2b_2+a_3b_3 & -a_2b_1 & -a_3b_1 \\
-a_1b_2 & a_3b_3+a_1b_1 & -a_3b_2 \\
-a_1b_3 & -a_2b_3 & a_1b_1+a_2b_2 
\ea \right) 
\right. \\ \left.
+y^a_A y^s_B \left( \ba{ccc}
-a_2b_2+a_3b_3 & -a_2b_1 & a_3b_1 \\
a_1b_2 & -a_3b_3+a_1b_1 & -a_3b_2 \\
-a_1b_3 & a_2b_3 & -a_1b_1+a_2b_2 
\ea \right)
\right. \\ \left.
+y^s_A y^a_B \left( \ba{ccc}
-a_2b_2+a_3b_3 & a_2b_1 & -a_3b_1 \\
-a_1b_2 & -a_3b_3+a_1b_1 & a_3b_2 \\
a_1b_3 & -a_2b_3 & -a_1b_1+a_2b_2 
\ea \right)
\right]\displaystyle{\frac{1}{M}}
\ea$
\\ \hline
 \end{tabular}
 \caption{The $A_4$ flavour structures arising from dim-5 operators,
 defined as in Eq.~(\ref{d5}).
In the cases A, A$'$, A$''$, C and D one may identify both the two Higgs multiplets,
$\Phi_A\equiv \Phi_B$, and the pair of messengers, $\Sigma \equiv \overline{\Sigma}$.
If these identifications are made, one has to take 
$y_A\equiv y_B$ (with possible superscripts understood).}
 \label{tab:fl5D}
\end{center}\end{table}

The $SO(10)$ assignments of $\Phi_{A,B}$ and $(\Sigma,\overline{\Sigma})$
determine the relative contribution to the mass matrices of the different sectors. 
In Table \ref{tab:ga5D} we classify all dim-5 operators involving representations 
with size  $\le 120$.
It is useful to recall that an $SO(10)$ adjoint Higgs multiplet $45_H$
can acquire VEVs in two independent directions. In the Table
we indicate with $b_1$ ($b_{24}$)
the VEV in the $SU(5)$ singlet (adjoint) direction.
Equivalently, one can write the VEV of $45_H$ as a combination 
of a VEV $b_3$ in the right-handed isospin
direction and a VEV $b_{15}$ in the $B-L$ direction,
by using
\beq
b_1=\frac 15 (b_3+3b_{15}) ~,~~~~~ b_{24}=\frac 15 (-b_3+2b_{15}) ~.
\label{PS}\eeq
The Clebsch-Gordan coefficients in Table \ref{tab:ga5D} can be derived e.g.
with the help of Refs.\cite{NS1,NS2,syed,AG,FIKMO}.

For a given $SO(10)\times A_4$ assignment of $\Phi_A$, $\Phi_B$, $\Sigma$ and
$\overline{\Sigma}$,
the contribution to the mass matrices is obtained 
by combining the corresponding rows of Tables \ref{tab:fl5D} and \ref{tab:ga5D}.

Operators with dimension larger than 5 can also correct significantly fermion masses,
in particular when the cutoff $M$ is not much larger than $M_{GUT}$. 
An $SO(10)\times A_4$ model which crucially relies on a dim-6 operator
was built in \cite{Morisi:2007ft}. In this model  the operator VIII of
Table \ref{tab:ga5D} is used, with $120_H$ replaced by a product $45_H 10_H$.

\begin{center}
\begin{table}[hbtp]
\begin{tabular}{|c|c|c|}
\hline
case & $SO(10)$ operator & mass matrices  \\
\hline
\hline 
IV & $(16_M 16_H)_{10} (16_M 16_H)_{10}$ &
$\begin{array} {l}  
M_d=     K a_{\overline{5}} b_1 + K^T a_1 b_{\overline 5} \\
M_e^T= K a_{\overline{5}} b_1 + K^T a_1 b_{\overline 5} \\
\end{array}$\\
\hline
V & $(16_M \overline{16}_H)_1 (16_M \overline{16}_H)_1$ & 
$\begin{array} {l}  
M_\nu = K a_5 b_1 + K^T a_1 b_5 \\
M_L = K_s a_5 b_5 \\
M_R = K_s a_1 b_1 \\
\end{array}$\\
\hline 
VI & $(16_M \overline{16}_H)_{45} (16_M \overline{16}_H)_{45}$ &
$\begin{array}{l} 
M_u = 8 K_s (a_5 b_1 + a_1 b_5) \\
M_\nu = 3 (K a_5 b_1 + K^T a_1 b_5) \\
M_L = -5 K_s a_5 b_5 \\
M_R = -5 K_s a_1 b_1 \\
\end{array}$\\
\hline
VII & $(16_M 10_H)_{16} (16_M 45_H)_{\overline{16}}$ & 
$\begin{array}{l} 
M_u = K a_5 (b_1 - 4 b_{24}) + K^T a_5 ( b_1 + b_{24}) \\
M_\nu = 5 K a_5 b_1 + K^T a_5 (-3 b_1 - 3 b_{24})  \\
M_d = K a_{\overline 5} (-3b_1 + 2 b_{24})
+ K^T a_{\overline 5} (b_1 + b_{24}) \\
M_e^T = K a_{\overline 5} (-3b_1 - 3 b_{24})
+ K^T a_{\overline 5} (b_1 + 6 b_{24}) \\
\end{array}$\\
\hline
VIII & $(16_M 120_H)_{16} (16_M 45_H)_{\overline{16}}$ & 
$\begin{array} {l} 
M_u =K a_{45}(b_1-4b_{24})-K^T a_{45}(b_1+b_{24})\\
M_\nu= 5 K a_{5}b_1 - K^T a_{5}(-3b_1-3b_{24})\\
M_d =K (a_{\overline{5}}+a_{\overline{45}})(-3b_1+2b_{24})
-K^T (a_{\overline{5}}+a_{\overline{45}})(b_1+b_{24})\\
M_e^T=K (a_{\overline{5}}-3 a_{\overline{45}})(-3b_1-3b_{24})
-K^T (a_{\overline{5}}- 3 a_{\overline{45}})(b_1+6b_{24})\\
 \end{array} $ \\
\hline
IX & $(16_M 16_H)_{120} (16_M 16_H)_{120}$ & 
$\ba{l}
M_d = K (a_{\overline{5}} b_1 + 2 a_1 b_{\overline{5}} )
+ K^T (a_1 b_{\overline{5}} +2 a_{\overline{5}} b_1)\\
M_e^T = K (a_{\overline{5}} b_1 + 2 a_1 b_{\overline{5}} )
+ K^T (a_1 b_{\overline{5}} +2 a_{\overline{5}} b_1)\\
\ea$
\\ \hline
\end{tabular}
\caption{
The contributions to the mass matrices from $SO(10)$-invariant dim-5 operators,
defined as in Eq.~(\ref{d5}). 
Here we introduced 
$K \equiv y_A y_B/M$ and $K_s\equiv (K+K^T)/2$. 
This list exhausts all possibilities with $\Phi_A$, $\Phi_B$,
$\Sigma$ and $\overline{\Sigma}$ in representations with  size $\le 120$.
Different VEVs of the same $SO(10)$ Higgs multiplet carry a subscript indicating the 
$SU(5)$ component they belong to.}
\label{tab:ga5D}
\end{table}
\end{center}

\vskip 20pt

\end{document}